\documentclass[apj, numberedappendix]{emulateapj}
\usepackage{epsfig}
\usepackage{graphicx}

\usepackage{amsfonts}
\usepackage{amsmath}
\usepackage{mathrsfs}
\usepackage{amssymb}
\usepackage{rotating}
\usepackage{float}

\usepackage{braket}
\usepackage{longtable}
\usepackage[para]{threeparttablex}
\usepackage{adjustbox}
\usepackage{afterpage}
\usepackage{bm}
\usepackage{setspace}

\usepackage{natbib}
\usepackage[pdfstartview=FitH,      
            breaklinks=true,        
            bookmarksopen=true,     
            bookmarksnumbered=true  
            ]{hyperref}

\usepackage{color}
\usepackage{ulem}
\usepackage{xspace}
\usepackage{cancel}
\definecolor{lgray}{gray}{0.85}

\newcommand{\lya} {Ly$\alpha$\xspace}
\newcommand{\lyb} {Ly$\beta$\xspace}
\newcommand{\lyc} {Ly$\gamma$\xspace}

\newcommand{\siiv} {\ion{Si}{4}\xspace}

\newcommand{\hii} {\ion{H}{2}\xspace}
\newcommand{\cii} {[\ion{C}{2}]\xspace}
\newcommand{\mgii} {\ion{Mg}{2}\xspace}
\newcommand{\siii} {\ion{Si}{2}\xspace}

\newcommand{\oi} {\ion{O}{1}\xspace}

\newcommand{\heii}{\ion{He}{2}\xspace}

\newcommand{\nv} {\ion{N}{5}\xspace}

\def\bea{\begin{eqnarray}}
\def\eea{\end{eqnarray}}
\newcommand*\diff{\mathop{}\!\mathrm{d}}

\begin{document}

\title{Implications of $z \sim 6$ Quasar Proximity Zones for the Epoch of Reionization and Quasar Lifetimes} 

\author{Anna-Christina Eilers\altaffilmark{1,2*}, Frederick B. Davies\altaffilmark{1}, Joseph F. Hennawi\altaffilmark{1,3}, J. Xavier Prochaska\altaffilmark{4}, Zarija Luki{\'c}\altaffilmark{5}, Chiara Mazzucchelli\altaffilmark{1,2}}
\altaffiltext{*}{email: eilers@mpia.de}
\altaffiltext{1}{Max-Planck-Institute for Astronomy, K\"onigstuhl 17, 69117 Heidelberg, Germany; eilers@mpia.de}
\altaffiltext{2}{International Max Planck Research School for Astronomy \& Cosmic Physics at the University of Heidelberg}
\altaffiltext{3}{Physics Department, University of California, Santa Barbara, CA 93106-9530, USA}
\altaffiltext{4}{Department of Astronomy and Astrophysics, University of California, Santa Cruz, CA 95064, USA}
\altaffiltext{5}{Lawrence Berkeley National Laboratory, CA 94720-8139, USA}

\slugcomment{Draft Version of \today}
\shortauthors{Eilers et al.}

\begin{abstract}
  We study quasar proximity zones in the redshift range $5.77 \leq z
  \leq 6.54$ by homogeneously analyzing $34$ medium resolution spectra,
  encompassing both archival and newly obtained data, and exploiting
  recently updated systemic redshift and magnitude
  measurements. Whereas previous studies found strong evolution of
  proximity zone sizes with redshift, and argued that this provides
  evidence for a rapidly evolving intergalactic medium (IGM) neutral
  fraction during reionization, we measure a much shallower trend
  $\propto(1+z)^{-1.44}$. We compare our measured proximity
  zone sizes to predictions from hydrodynamical simulations
  post-processed with one-dimensional radiative transfer, and find
  good agreement between observations and theory irrespective of the
  ionization state of the ambient IGM. This insensitivity to IGM ionization
  state has been previously noted, and results from the fact 
  that the definition of proximity zone size as the first drop of the
  smoothed quasar spectrum below the $10\%$ flux transmission level
  probes locations where the ionizing radiation from the quasar is an
  order of magnitude larger than the expected ultraviolet ionizing
  background that sets the neutral fraction of the IGM. Our analysis
  also uncovered three objects with exceptionally small proximity
  zones (two have $R_p < 1$~proper Mpc), which constitute outliers from the
  observed distribution and are challenging to explain with our radiative
  transfer simulations. We consider various explanations for their origin,
  such as strong absorption line systems associated with the quasar
  or patchy reionization, but find that the most compelling scenario
  is that these quasars have been shining for $\lesssim 10^5$~yr. 
\end{abstract}
    
\keywords{intergalactic medium --- epoch of reionization, dark ages --- methods: data analysis --- quasars: absorption lines} 

\maketitle

\section{Introduction}

A major goal of observational cosmology is to understand the epoch of
reionization, when the universe transitioned from the
cosmic ``dark ages" following recombination, into the ionized universe
we observe today.  Despite much progress in the last decade, there are
still many open questions regarding the exact timing and the
morphology of the reionization process. Studies of the evolution of
the (Lyman-$\alpha$) \lya absorption features in the spectra of
distant quasars are one of the key observational probes of this
epoch. Both the steep rise in the \lya optical depth of the
intergalactic medium (IGM) with redshift for $z\gtrsim 5.5$, as well
as the increased scatter in the measurements, suggest a qualitative
change in the state of the intergalactic medium (IGM), likely
resulting from a rapid rise in the volume averaged neutral fraction
which is expected to occur during the end stages of reionization
\citep{{Fan2006}, {Becker2015}}. 
Indeed, the absence of large
Gunn-Peterson (GP) troughs \citep{GunnPeterson1965} 
in spectra at $z\lesssim 5.5$ indicates the
epoch of hydrogen reionization must be completed at that time
\citep[][]{McGreer2015}.
However, constraining the neutral hydrogen fraction of the IGM at $z\gtrsim
6$ with quasar absorption spectroscopy 
has proven extremely difficult, because Lyman series transitions
are overly sensitive, and saturate already for 
volume averaged neutral hydrogen
fractions $\langle f_{\rm HI}\rangle \gtrsim 10^{-4}$.

In this paper we focus on a different technique that has been applied in previous work to constrain the neutral gas fraction at
$z\sim 6$. Near the end of the cosmic reionization epoch, luminous
quasars exhibit a region of enhanced transmission immediately blueward of the \lya emission line in
the so-called quasar proximity zone, before the onset of near complete Gunn-Peterson absorption. 
This enhanced transmission is caused by the radiation from the quasar itself, which ionizes the
surrounding IGM \citep[see e.g.][]{{MadauRees2000},
  {CenHaiman2000}, {HaimanCen2001}, {Wyithe2005}, {BoltonHaehnelt2007a},
    {BoltonHaehnelt2007b},{Lidz2007},
  {Bolton2011}, {Keating2015}}. 
It has been argued that the evolution of the 
proximity zone sizes with redshift can constrain
the late stages of the reionization epoch \citep{Fan2006, Carilli2010, Venemans2015}. 

\citet{HaimanCen2001} showed that in a very simplistic picture of
reionization in which isolated ionized \hii regions expand into a neutral uniform 
ambient IGM, the (proper)
size $R_{\rm ion}$ of the ionized 
region around the quasar depends on the 
neutral
hydrogen fraction $f_{\rm HI}$, the rate at which ionizing photons are emitted $\dot{N}_{\gamma}$, 
and the quasar age $t_{\rm Q}$:
\begin{align}
R_{\rm ion}\approx \left(\frac{3\dot{N}_{\gamma}t_{\rm Q}}{4\pi n_{\rm H}f_{\rm HI}}\right)^{1/3}, \label{eq:rp}
\end{align}
where $n_{\rm H}$ is the hydrogen number density. 
In eqn.~(\ref{eq:rp}) 
recombinations of the ionized gas inside the \hii region are neglected, 
which should be unimportant on the timescale that the quasar turned on, because the recombination timescale of hydrogen is comparable to the Hubble time. 
However, there are a number of other effects that are not taken into
consideration in this relation, such as overlapping ionized \hii
regions, large-scale structure effects, and pre-ionization by local
galaxies or the clumpiness of the IGM \citep[see also][]{{Fan2006},
  {BoltonHaehnelt2007a}, {BoltonHaehnelt2007b}, {Lidz2007},
  {Maselli2007}, {Maselli2009}, {Khrykin2016}}.
  
While eqn.~(\ref{eq:rp}) provides a reasonable description of the expansion rate of an \hii ionization front embedded
in a neutral IGM, it does not predict the distribution of residual neutral hydrogen within the ionized bubble. 
However, the \textit{observed} size of the proximity zone $R_p$ will depend sensitively on the fraction and distribution of the residual neutral hydrogen, because this can cause saturated absorption well before the ionization front is reached  \citep{BoltonHaehnelt2007a}. 
Thus the extent of the ionized \hii bubble around the quasar can be significantly larger than the observed proximity zone, which is defined to end where the transmitted flux drops below the $10\%$ level \citep[see e.g.][]{Fan2006, Carilli2010} and hence does not necessarily probe the location of the ionization front. 

Furthermore \citet{BoltonHaehnelt2007a} showed that a
quasar embedded in a highly ionized IGM can produce a proximity zone
that appears qualitatively similar to that of one in a neutral ambient IGM. 
They show that the observed size of the proximity zone of a quasar within an already ionized surrounding IGM is independent of the neutral gas fraction of the ambient IGM and as such, may not 
provide insights into the evolution of the neutral hydrogen fraction during the epoch of reionization at all. 

However, previous observational studies of the sizes of quasar proximity zones found evidence for a steep decrease in proximity zone size with increasing redshift within the redshift range of $5.7\leq z\leq 6.4$ \citep{Fan2006, Willott2007, Willott2010, Carilli2010}. Although there is large scatter in the observations, this has been interpreted as a strong evolution of the neutral gas fraction of the IGM assuming that the observed proximity zones trace the extent of the ionized \hii region presented in eqn.~(\ref{eq:rp}), i.e. $R_p\approx R_{\rm ion}$. 
With the discovery of the current highest redshift quasar $\rm ULAS J1120+0641$ \citep{Mortlock2011} and the analysis of its proximity zone \citep{Bolton2011, Simcoe2012, Venemans2015, Bosman2015},
this steep decrease has become somewhat shallower, but still indicates a strong evolution in proximity zone size with
redshift. 
However, the measurements of the proximity zone sizes contain a number of uncertainties, most importantly the
large uncertainties in the systemic redshifts of the quasars and the resulting ambiguity in the
beginning of the proximity zone.

The extent of the \hii region around the quasar additionally depends on the age of the quasar (eqn.~(\ref{eq:rp})). Even if the IGM is highly ionized prior to the quasar turning on, the gas in the ionized region responds on a finite timescale \citep{Khrykin2016} and therefore also the observed proximity zone sizes depend on the quasar ages. 
Thus, assumptions about the quasar ages are required to interpret
quasar proximity zones, but the age remains uncertain by several orders
of magnitude \citep{Martini2004}.

To be more precise, we will distinguish between
several different timescales that govern the duration of quasar
activity.  The \textit{duty cycle} $t_{\rm dc}$ refers to the total
time that galaxies shine as active quasars integrated over the age of
the universe. However, quasar activity could be episodic, and
we refer to the  duration $t_{\rm episodic}$ of a single emission episode 
as the \textit{episodic lifetime}. The sizes of quasar
proximity zones actually depend on the quasar \textit{age}.
If we denote by $t=0$ the time at which
light emitted by a quasar just reaches our telescopes on Earth,
then the quasar age, which we will henceforth refer to as
$t_{\rm Q}$, is defined such that the quasar actually
turned on at a time $-t_{\rm Q}$ in the past. Strictly speaking the age $t_{\rm Q}$ is a lower limit on the quasar episodic lifetime $t_{\rm episodic}$, because the quasar episode might indeed continue, which we can only record on Earth if we could conduct observations in the future.

The quasar duty cycle can be constrained by
comparing the number density of quasars to their host dark matter halo
abundance inferred from their clustering strength
\citep{HaimanHui2001, MartiniWeinberg2001, Martini2004, WhiteMartiniCohn2008, 
  Furlanetto2011}. 
But to date this method has yielded only weak constraints on
$t_{\rm dc} \sim 10^6-10^9$~yr owing to uncertainties in how
quasars populate dark
matter halos \citep{Shen2009, White2012, ConroyWhite2013, Cen2015}. An upper limit on the
duty cycle of quasars, $t_{\rm dc} < 10^9$~yr, is set by the
observed evolution of the quasar luminosity function, since the whole
quasar population rises and falls over roughly this timescale
\citep[e.g.][]{Osmer1998}.

Moreover, these constraints on the duty cycle of quasars do not give insights into the duration of the individual accretion episodes of the quasar activity. 
A population of quasars emitting $1000$ individual bursts each with a lifetime of $t_{\rm episodic}\sim10^5$~yr, would be indistinguishable from quasars with steady continuous emission for $t_{\rm episodic}\sim10^8$~yr. A lower limit on the individual quasar bursts of $t_{\rm episodic}\sim 3\times 10^4$~yr is based on the argument that quasars need to maintain their ionizing luminosity long enough to explain the observed
proximity zones in the \lya forest \citep[e.g.][]{Bajtlik1988, Khrykin2016}. 

It has been argued, that in order to grow the observed sizes of SMBHs,
i.e. $M_{\rm SMBH}\sim 10^9-10^{10} M_{\sun}$ at $z\sim 6-7$
\citep{Mortlock2011, Venemans2013, DeRosa2014, Wu2015}, the quasars
require very massive initial seeds and need to accrete matter over
timescales comparable to the age of the Universe at these high
redshifts \citep{HopkinsHernquist2009, Volonteri2010, Volonteri2012}. It follows that
the duty cycle of these quasars needs to be of the order of the Hubble
time. If quasar activity is episodic, the episodic bursts need to be
very long or the quiescent time in between the bursts needs to be
short for the SMBHs to grow to their observed sizes.

The dozens of $z\sim 6$ quasars that have been uncovered over the past
decade \citep[see][for a recent compilation]{Banados2016} from wide field surveys
results in many new quasars for proximity zone measurements.  The statistical power
of these data alone motivates revisiting this type of analysis to
further understand the resulting constraints on reionization and the
quasar emission timescales.  In this paper we re-investigate the
evolution of the quasar proximity zone sizes at $z\geq 5.77$ with a
homogeneous analysis of a significantly enlarged 
sample of $31$ quasar spectra with higher quality data, including updated
and more precise redshift measurements from CO and \cii line observations,
and consistently measured absolute magnitudes. We then compare our analysis
with state-of-the-art radiative transfer simulations in order to better understand and interpret our measurements. 

This paper is structured as follows: we describe our data set and the reduction pipeline in \S~\ref{sec:data}.
In \S~\ref{sec:methods} we specify the methods we use for the continuum normalization and the measurement of the proximity zone size of each quasar. In \S~\ref{sec:sims} we describe a suite of radiative transfer simulations that we compare to our observations. We show our measurements of the proximity zone sizes in \S~\ref{sec:z_evolution} and discuss their evolution with redshift. We highlight
three exceptionally small proximity zones 
in our sample in \S~\ref{sec:small_zones}, and discuss possible
explanations for their origin. We summarize and conclude in
\S~\ref{sec:summary}.

\section{High Redshift Quasar Sample}\label{sec:data}

Our initial data set consists of $34$ 
quasar spectra at $5.77\leq z_{\rm em} \leq 6.54$, 
observed at optical wavelengths ($4000$~{\AA} - $10000$~{\AA}) with the Echellette Spectrograph and Imager \citep[ESI;][]{ESI} at the Keck II Telescope in the years 2001 to 2016. 
We collected data from the Keck Observatory Archive\footnote{\url{https://koa.ipac.caltech.edu/cgi-bin/KOA/nph-KOAlogin}} and supplemented it with our own observations. 
The slit widths used range from $0.75"-1.0"$, giving a resolution of $R\approx 4000-5400$. 
The exposure times vary from $0.3$~h $\lesssim t_{\rm exp}\lesssim 25$~h resulting in median
signal-to-noise ratios in the quasar continuum at rest-frame wavelength of $1250${\AA}\,-$1280${\AA}\, in the
range of $2\lesssim \rm S/N \lesssim 108$ per pixel. 

We obtained spectra of the four quasars
$\rm PSO J0226+0302$\footnote{also known as PSO J036.5078+03.0498 \citep{Venemans2015}.}, $\rm PSO J0402+2452$\footnote{also known as PSO J060.5529+24.8567 \citep{Banados2016}. }, $\rm SDSS J0100+2802$, and $\rm SDSS J1137+3549$
with ESI on on January 11th and 12th, 2016. We used a $1"$ slit and exposure times varied between $1-3.25$~hours. 

For one object $\rm CFHQS J0227-0605$ of interest because of its small
proximity zone (see \S~\ref{sec:small_zones}), we obtained an additional spectrum
with the Low Resolution Imaging Spectrometer \citep[LRIS;][]{LRIS} on
the Keck I Telescope with an exposure time of $3600$~s. The observations were conducted on September
16th, 2016 using the $600/7500$ grating and a slit width of $1"$,
resulting in a resolution of $R\approx 1800$. Note that we only use this LRIS spectrum
for follow-up analysis, i.e. searching for
metal absorption lines. For the main analysis of the proximity zone of
this quasar we use the ESI spectrum, in order to be consistent with
the remaining data sample.

The details of all of the observations can be found in Table~\ref{tab:observations}.

\LongTables
\begin{deluxetable*}{@{\extracolsep{\fill}}lcclrr@{}}
\tablewidth{0.99\textwidth}
\tablecaption{Overview of the observations of the $34$ quasars in our data sample. \label{tab:observations}}
\startdata
\tablehead{\colhead{object} & \colhead{RA (J2000)}  &  \colhead{DEC (J2000)} & \colhead{PI} & \colhead{observation date} & \colhead{exposure time}}
SDSS J0002+2550 &  $00^\mathrm{h}02^\mathrm{m}39\fs39$ & $+25\degr50\arcmin34\farcs96$ & Kakazu & Nov. 2004& $5400$~s\\
 &  &  & Cowie & Aug. 2005 & $16300$~s\\
SDSS J0005-0006 &  $00^\mathrm{h}05^\mathrm{m}52\fs34$ & $-00\degr06\arcmin55\farcs80$ & Becker & Dec. 2002 & $1200$~s\\
 &  &  & Sargent & Oct. 2010 & $15000$~s\\
CFHQS J0050+3445 &  $00^\mathrm{h}55^\mathrm{m}02\fs91$ & $+34\degr45\arcmin21\farcs65$ & Willott & Sep. 2008& $6250$~s\\
 &  &  & Sargent & Oct. 2010 & $6000$~s\\
 &  &  & Sargent & Oct. 2012 & $9000$~s\\
SDSS J0100+2802 &  $01^\mathrm{h}00^\mathrm{m}13\fs02$ & $+28\degr02\arcmin25\farcs92$ & White & Jan. 2016& $3600$~s\\
ULAS J0148+0600 &  $01^\mathrm{h}48^\mathrm{m}37\fs64$ & $+06\degr00\arcmin20\farcs06$ & Sargent & Oct. 2010 & $11200$~s\\
ULAS J0203+0012 &  $02^\mathrm{h}03^\mathrm{m}32\fs38$ & $+00\degr12\arcmin29\farcs27$ & Sargent & Oct. 2010 & $6600$~s\\
CFHQS J0210-0456 &  $02^\mathrm{h}10^\mathrm{m}13\fs19$ & $-04\degr56\arcmin20\farcs90$ & Sargent & Oct. 2010 & $6000$~s\\
PSO J0226+0302 & $02^\mathrm{h}26^\mathrm{m}01\fs87$ & $+03\degr02\arcmin59\farcs42$ & White & Jan. 2016& $11700$~s\\
CFHQS J0227-0605 &  $02^\mathrm{h}27^\mathrm{m}43\fs29$ & $-06\degr05\arcmin30\farcs20$ & Willott & Sep. 2008 & $10540$~s\\
SDSS J0303-0019 &  $03^\mathrm{h}03^\mathrm{m}31\fs40$ & $-00\degr19\arcmin12\farcs90$ & Sargent & Oct. 2010& $6000$~s\\
SDSS J0353+0104 &  $03^\mathrm{h}53^\mathrm{m}49\fs73$ & $+01\degr04\arcmin04\farcs66$ & Sargent & Oct. 2010 & $13200$~s\\
 &  &  & Becker & Jan. 2006& $3600$~s\\
PSO J0402+2452 &  $04^\mathrm{h}02^\mathrm{m}12\fs69$ & $+24\degr51\arcmin24\farcs43$ & White & Jan. 2016& $10800$~s\\
SDSS J0818+1723 &  $08^\mathrm{h}18^\mathrm{m}27\fs40$ & $+17\degr22\arcmin52\farcs01$ & Becker & Apr. 2005& $2400$~s\\
SDSS J0836+0054 &  $08^\mathrm{h}36^\mathrm{m}43\fs86$ & $+00\degr54\arcmin53\farcs26$ & Becker & Mar. 2001& $1731$~s\\
 &  &  & Madau & Feb. 2002 & $18900$~s\\
 &  &  & Cowie & Feb. 2002 & $13540$~s\\
 &  &  & Djorgovski & Mar. 2002 & $	8923$~s\\
 &  &  & Kulkarni & Jan. 2003 & $5400$~s\\
 &  &  & Cowie & Jan. 2003 & $3600$~s\\
 &  &  & Cowie & Jan. 2004 & $10800$~s\\
 &  &  & Kakazu & Nov. 2004 & $10200$~s\\
 &  &  & Djorgovski & Dec. 2004 & $7200$~s\\
SDSS J0840+5624 &  $08^\mathrm{h}40^\mathrm{m}35\fs30$ & $+56\degr24\arcmin20\farcs22$ & Djorgovski & Dec. 2004& $10800$~s\\
 &  &  & Becker & Mar. 2006& $1200$~s\\
SDSS J0842+1218 &  $08^\mathrm{h}42^\mathrm{m}29\fs43$ & $+12\degr18\arcmin50\farcs58$ & Becker & Mar. 2006& $2400$~s\\
SDSS J0927+2001 &  $09^\mathrm{h}27^\mathrm{m}21\fs82$ & $+20\degr01\arcmin23\farcs64$ & Becker & Mar. 2006& $1200$~s\\
SDSS J1030+0524 &  $10^\mathrm{h}30^\mathrm{m}27\fs11$ & $+05\degr24\arcmin55\farcs06$ & Becker & May 2001& $1800$~s\\
 &  &  & Becker & Jan. 2002 & $12000$~s\\
 &  &  & Madau & Feb. 2002 & $7339$~s\\
 &  &  & Cowie & Feb. 2002 & $16200$~s\\
 &  &  & Djorgovski & Mar. 2002 & $16200$~s\\
 &  &  & Cowie & Jan. 2003 & $5400$~s\\
SDSS J1048+4637 &  $10^\mathrm{h}48^\mathrm{m}45\fs07$ & $+46\degr37\arcmin18\farcs55$ & Becker & Dec. 2002& $3600$~s\\
 &  &  & Cowie & Jan. 2003 & $18000$~s\\
 &  &  & Cowie & Mar. 2003 & $5400$~s\\
 &  &  & Sanchez & Jun. 2003 & $5400$~s\\
 &  &  & Djorgovski & Dec. 2004 & $6600$~s\\
SDSS J1137+3549  &  $11^\mathrm{h}37^\mathrm{m}17\fs73$ & $+35\degr49\arcmin56\farcs85$ & White & Jan. 2016& $7800$~s\\
 &  &  & Becker & Jan. 2005 & $2400$~s\\
SDSS J1148+5251 &  $11^\mathrm{h}48^\mathrm{m}16\fs65$ & $+52\degr51\arcmin50\farcs39$ & Becker & May 2002 & $8100$~s\\*
 &  &  & Becker & Dec. 2002 & $22800$~s\\
 &  &  & Djorgovski & Dec. 2002 & $11400$~s\\
 &  &  & Kulkarni & Jan. 2003 & $3600$~s\\
 &  &  & Cowie & Jan. 2003 & $16200$~s\\
 &  &  & Becker & Feb. 2003 & $16800$~s\\
 &  &  & Cowie & Jan. 2004 & $11700$~s\\
SDSS J1250+3130 &  $12^\mathrm{h}50^\mathrm{m}51\fs93$ & $+31\degr30\arcmin21\farcs90$ & Becker & Jan. 2005& $3600$~s\\
SDSS J1306+0359 &  $13^\mathrm{h}06^\mathrm{m}08\fs27$ & $+03\degr59\arcmin26\farcs36$ & Becker & May 2001& $900$~s\\ 
 &  &  & Cowie & Feb. 2002 & $15300$~s\\
 &  &  & Djorgovski & Mar. 2002 & $12600$~s\\
 &  &  & Cowie & Mar. 2003 & $9000$~s\\
 &  &  & Cowie & Mar. 2004 & $12600$~s\\
ULAS J1319+0950 & $13^\mathrm{h}19^\mathrm{m}11\fs30$ & $+09\degr50\arcmin51\farcs52$ & Steidel & Mar. 2008& $2400$~s\\
SDSS J1335+3533 &  $13^\mathrm{h}35^\mathrm{m}50\fs81$ & $+35\degr33\arcmin15\farcs82$ & Becker & Mar. 2006& $1200$~s\\
SDSS J1411+1217 &  $14^\mathrm{h}11^\mathrm{m}11\fs29$ & $+12\degr17\arcmin37\farcs28$ & Djorgovski & Apr. 2004& $10200$~s\\
 &  &  & Becker & Jan. 2005 & $3600$~s\\
 &  &  & Cowie & May 2005 & $10800$~s\\
 &  &  & Cowie & Mar. 2008 & $12600$~s\\
SDSS J1602+4228 & $16^\mathrm{h}02^\mathrm{m}53\fs98$ & $+42\degr28\arcmin24\farcs94$ & Cowie & May. 2005& $10620$~s\\
 &  &  & Cowie & Aug. 2005 & $3600$~s\\
SDSS J1623+3112 &  $16^\mathrm{h}23^\mathrm{m}31\fs81$ & $+31\degr12\arcmin00\farcs53$ & Becker & Jan. 2004& $	3600$~s\\
SDSS J1630+4012 &  $16^\mathrm{h}30^\mathrm{m}33\fs90$ & $+40\degr12\arcmin09\farcs69$ & Sanchez & Jun. 2003& $17700$~s\\
 &  &  & Sargent & Oct. 2010 & $2000$~s\\
CFHQS J1641+3755 &  $16^\mathrm{h}41^\mathrm{m}21\fs73$ & $+37\degr55\arcmin20\farcs15$ & Willott & Oct. 2007& $2400$~s\\
SDSS J2054-0005 &  $20^\mathrm{h}54^\mathrm{m}06\fs49$ & $-00\degr05\arcmin14\farcs80$ & Sargent & Oct. 2010 & $12000$~s\\
CFHQS J2229+1457 &  $22^\mathrm{h}29^\mathrm{m}01\fs65$ & $+14\degr57\arcmin09\farcs00$ & Willott & Sep. 2008 & $3600$~s\\
 &  &  & Prochaska\tablenotemark{*} & Sep. 2016 & $3600$~s\\
SDSS J2315-0023 &  $23^\mathrm{h}15^\mathrm{m}46\fs57$ & $-00\degr23\arcmin58\farcs10$ & Becker & Jan. 2006& $1200$~s\\
 &  &  & Sargent & Oct. 2010 & $27000$~s\\
CFHQS J2329-0301 &  $23^\mathrm{h}29^\mathrm{m}08\fs28$ & $-03\degr01\arcmin58\farcs80$ & Willott & Oct. 2007& $10800$~s\\
& & & Willott & Sep. 2008& $14400$~s
\enddata
\tablecomments{Columns show the object name, its coordinates, the principal investigator of the different observing runs, the observation date and total exposure time of the object in this run.}
\tablenotetext{*}{spectrum taken with LRIS}
\end{deluxetable*}

\subsection{Data Reduction}

We reduce all spectra uniformly using the ESIRedux pipeline\footnote{\url{http://www2.keck.hawaii.edu/inst/esi/ESIRedux/}} developed as part of the XIDL\footnote{\url{http://www.ucolick.org/~xavier/IDL/}} suite of astronomical routines in the Interactive Data Language (IDL). 
This pipeline employs standard data reduction techniques which can be
summarized as follows: Images are overscan subtracted, flat fielded using
a normalized flat field image, and then wavelength calibrated using a wavelength image constructed
from afternoon arc lamp calibration images. Objects are identified in the science frames, and then
background subtracted using $B$-spline fits \citep{Kelson2003, Bernstein2015} to object free regions of the slit. Object
profiles are also fit with $B$-splines, and optimal extraction 
is performed on the sky-subtracted
frames. One-dimensional spectra of overlapping echelle orders are combined to produce a final spectrum for each
exposure, and individual exposures are co-added into our final one-dimensional spectra. See \citep{Bochanski2009}
for a more detailed description of the algorithms used.

We further optimized the XIDL ESI pipeline to improve the data
reduction for high redshift quasars. The main improvement was to
remove the fringing pattern from reddest orders by differencing two
images (ideally taken during the same run) with similar exposure
times, analogous to the standard difference imaging techniques
performed for near-infrared observations. However, this procedure
only works on dithered exposures for which the trace of the science
object lands at different spatial locations on the slit.  Since not
every observer dithered their object along the slit it was not
possible for us to apply this procedure to $\approx 10\%$ of the
exposures that we took from the archive.

We also co-added exposures from different observing runs taken by different PIs, resulting in higher $\rm S/N$ data for some quasars than previous analyses of these objects. 
We weight each one-dimensional spectrum by its squared signal-to-noise ratio ($\rm S/N^2$)
that was determined in the quasar continuum region of each spectrum,
i.e. at wavelengths longer than the \lya emission line.  In this way
spectral regions with low or no transmitted flux, which are common in
high redshift quasar spectra, are weighted by the same $\rm S/N$ ratio of regions with more transmitted flux.

\subsection{Quasar Properties}

Determining precise redshifts for quasars is very challenging due to the broad widths of emission lines, Gunn-Peterson absorption, and offsets between different ionization lines \citep{Gaskell1982, TytlerFan1992, VandenBerk2001, Richards2002a, Shen2016}. 
Most quasars show strong internal motions and winds, which displace many of the emission lines, such as the \lya line or far-ultraviolet (far-UV) lines, far from the systemic redshift of the host galaxy.
Thus the most precise determination of the location of the quasar are low ionization 
lines such as \mgii lines or, even better, emission lines from the molecular gas reservoir of the host galaxy itself, such as CO or \cii lines.

\begin{deluxetable*}{@{\extracolsep{\fill}}llrlrcrcc@{}}
\tabletypesize{\footnotesize}
\tablewidth{0.99\textwidth}
\tablecaption{Overview of our data sample and the measurements of the proximity zone sizes. \label{tab:overview}}
\startdata
\tablehead{\colhead{object} & \colhead{$z$} & \colhead{$\Delta z~\rm [km/s]$} & \colhead{redshift line} & \colhead{Ref.$^{\rm a}$} & \colhead{$M_{\rm 1450}$} & \colhead{$\rm S/N^b$} & \colhead{$R_p$ [pMpc]} & \colhead{$R_{p, \rm corr}$ [pMpc]}}
PSO J0226+0302 & $6.5412$ & $100$ &\cii & 13 &$-27.33$& $10$ &$3.64\pm0.13$ & $3.20\pm0.11$ \\
CFHQS J0210-0456 & $6.4323$ & $100$ &\cii & 12  &$-24.53$& $2$ &$1.32\pm0.13$ & $3.47\pm0.34$ \\
SDSS J1148+5251 & $6.4189$ & $100$ &\cii & 15 &$-27.62$& $34$&$4.58\pm0.13$ & $3.59\pm0.10$ \\
CFHQS J2329-0301 & $6.417$ & $270$ &\mgii & 9 &$-25.25$& $2$&$2.45\pm0.35$ & $4.86\pm0.70$ \\
SDSS J0100+2802 & $6.3258$ & $100$ &\cii & 17 &$-29.14$& $41$&$7.12\pm0.13$ & $3.09\pm0.06$ \\
SDSS J1030+0524 & $6.309$& $270$ &\mgii & 3 &$-26.99$& $26$&$5.93\pm0.36$ & $5.95\pm0.36$ \\
SDSS J1623+3112 & $6.2572$& $100$ &\cii & 18 &$-26.55$& $9$&$5.05\pm0.14$ & $6.03\pm0.16$ \\
CFHQS J0050+3445 & $6.253$ & $270$ &\mgii & 9 &$-26.70$&$18$ &$4.09\pm0.37$ & $4.60\pm0.41$ \\
CFHQS J0227-0605 & $6.20$ & $1000$ &\lya  & 6 &$-25.28$& $3$ &$1.60\pm1.37$ & $3.15\pm2.69$ \\
PSO J0402+2452 & $6.18$ & $1000$ &\lya  & 16 &$-26.95$& $13$&$4.17\pm1.38$ & $4.26\pm1.40$ \\
CFHQS J2229+1457 & $6.1517$ & $100$ &\cii & 15 &$-24.78$& $2^{\rm c}$ &$0.45\pm0.14$ & $1.07\pm0.33$ \\
SDSS J1250+3130 & $6.15$ &$1000$ &\lya-\oi-\siiv  & 2 &$-26.53$& $8$ &$6.59\pm1.38$ & $7.93\pm1.66$ \\
ULAS J1319+0950 & $6.1330$& $100$ &\cii & 11 &$-27.05$& $10$ &$3.84\pm0.14$ & $3.77\pm0.14$ \\
SDSS J2315-0023 & $6.117$ & $1000$ &\lya  & 5 &$-25.66$& $15$ &$3.70\pm1.39$ & $6.26\pm2.36$ \\
SDSS J1602+4228 & $6.09$ & $1000$ &\lya-\oi  & 1 &$-26.94$& $21$ &$7.11\pm1.40$ & $7.28\pm1.43$ \\
SDSS J0303-0019 & $6.078$ & $270$&\mgii & 7 &$-25.56$& $2$ &$2.21\pm0.38$ & $3.88\pm0.67$ \\
SDSS J0842+1218 & $6.069$&  $270$&\mgii  & 10 &$-26.91$& $10$ &$6.47\pm0.38$ & $6.71\pm0.39$ \\
SDSS J1630+4012 & $6.065$  & $270$&\mgii & 7 &$-26.19$& $15$ &$4.80\pm0.38$ & $6.59\pm0.52$ \\
CFHQS J1641+3755 & $6.047$  & $270$&\mgii & 9 &$-25.67$& $4$ &$3.98\pm0.38$ & $6.71\pm0.64$ \\
SDSS J2054-0005 & $6.0391$ & $100$ &\cii & 11 &$-26.21$& $17$ &$3.17\pm0.14$ & $4.32\pm0.19$ \\
SDSS J1137+3549 & $6.03$ & $1000$ & \lya-\oi-\siiv   & 2 &$-27.36$& $24$ &$6.98\pm1.42$ & $6.06\pm1.23$ \\
SDSS J0818+1723 & $6.02$ & $1000$ & \lya  & 2 &$-27.52$& $8$ &$5.89\pm1.42$ & $4.80\pm1.16$ \\
SDSS J1306+0359 & $6.016$ & $270$ &\mgii & 4 &$-26.81$& $41$ &$5.39\pm0.38$ & $5.80\pm0.41$ \\
ULAS J0148+0600 & $5.98$ & $270$ &\mgii & 14  &$-27.39$& $30$ &$6.03\pm0.39$ & $5.18\pm0.33$ \\
SDSS J1411+1217 & $5.904$& $270$ &\mgii & 4 &$-26.69$& $32$ &$4.60\pm0.39$ & $5.19\pm0.44$ \\
SDSS J1335+3533 & $5.9012$ & $100$ &CO  & 8 &$-26.67$& $7$ &$0.78\pm0.15$ & $0.89\pm0.17$ \\
SDSS J0840+5624$^{\rm d}$ & $5.8441$ &$100$ &CO & 8 &$-27.24$& $28$ &$0.88\pm0.15$ & $0.80\pm0.13$ \\
SDSS J0005-0006 & $5.844$ & $270$&\mgii & 10 &$-25.73$& $15$ &$2.87\pm0.40$ & $4.73\pm0.66$ \\
SDSS J0002+2550 & $5.82$ & $1000$ &\lya-\oi  & 1 &$-27.31$& $62$&$5.43\pm1.49$ & $4.81\pm1.31$ \\
SDSS J0836+0054 & $5.810$ &$270$ &\mgii & 4 &$-27.75$& $120$ &$5.06\pm0.40$ & $3.77\pm0.30$ \\
SDSS J0927+2001 & $5.7722$ &$100$ &CO  & 8 &$-26.76$& $7$ &$4.68\pm0.15$ & $5.14\pm0.16$
\enddata
\tablecomments{The columns show the object name, the redshift of the quasar and the redshift uncertainty, the lines measured to determine the redshift and the reference therefor, the quasar's magnitude $M_{\rm 1450}$ and the $\rm S/N$ of the its spectrum and our measurements for the proximity zones, uncorrected and luminosity corrected. }
\tablenotetext{a}{Reference for redshift. 1: \citet{Fan2004}, 2: \citet{Fan2006_discovery}, 3: \citet{Jiang2007}, 4: \citet{Kurk2007}, 5: \citet{Jiang2008}, 6: \citet{Willott2009}, 7: \citet{Carilli2010}, 8: \citet{Wang2010}, 9: \citet{Willott2010}, 10: \citet{DeRosa2011}, 11: \citet{Wang2013}, 12: \citet{Willott2013}, 13: \citet{Banados2015}, 14: \citet{Becker2015}, 15: \citet{Willott2015}, 16: \citet{Banados2016}, 17: \citet{Wang2016}, 18: private communication with R. Wang. }
\tablenotetext{b}{Median $\rm S/N$ per pixel; estimated between $1250$~{\AA}$\leq\lambda_{\rm rest}\leq1280$~{\AA}. }
\tablenotetext{c}{The LRIS spectrum we have for this object has $\rm S/N=7$. }
\tablenotetext{d}{Excluded from our analysis due to associated absorbers (see Appendix~\ref{sec:J0840}). }
\end{deluxetable*}

We account for uncertainties in the systemic redshift of each quasar
depending on the measured emission lines to determine their
redshifts. For redshifts determined from the detection of a \cii line
at $158\,\rm\mu$m or a CO line we assign a redshift error of $\Delta
v=100$~km/s. For quasars with a redshift measurement from a \mgii line
we assume a redshift error of $\Delta v=270$~km/s, in order to account
for the dispersion between the redshift of the \mgii line and the redshift of the host galaxy
\citep[e.g.][]{Richards2002a, Hennawi2006, Shen2012, Venemans2016, Shen2016}, 
and for the remaining quasars
for which the redshift was determined by the \lya line or far-UV
lines, which usually suffer from large velocity offsets, we assume a redshift error of $\Delta v=1000$~km/s. Note
that these uncertainties are chosen to be very conservative
\citep{Shen2016}, because the redshift measurements are taken from the
literature, measured by various authors possibly using different data
and methods. 
For a quasar at $z=6$ these offsets in velocity result in a distance uncertainty of $\Delta R\approx 0.14$~proper Mpc (pMpc) for quasars with redshift measurements from CO or \cii lines, $\Delta R\approx 0.39$~pMpc for quasars with redshift measurements from their \mgii line, and $\Delta R\approx 1.43$~pMpc for
other rest-frame UV lines.

We take the absolute magnitudes $M_{\rm 1450}$ defined at $\lambda_{\rm rest}=1450$~{\AA}
in the rest-frame from \citet{Banados2016}, who determined $M_{\rm 1450}$ for all sources in a consistent way. They assume a power law continuum slope $\alpha_{\nu}=-0.3$ and require the extrapolated $y_{\rm P1}$- or $J$-band magnitudes (at $\lambda_{\rm eff}=9627.7$~{\AA} or $\lambda_{\rm eff}=12444.0$~{\AA}, respectively) to be consistent with their measurements, 
since most optical quasar spectra at these high redshifts have limited wavelength coverage at $\lambda_{\rm rest}=1450${\AA}. Table \ref{tab:overview} summarizes the properties of all quasars in our sample.

Note that we exclude three quasars ($\rm SDSS J1048+4637$, $\rm SDSS J0353+0104$, and $\rm ULAS J0203+0012$) from this analysis, because of their broad absorption line (BAL) features, which can contaminate the proximity zone and make it difficult to determine a precise redshift or the continuum level of these quasars. 
For a similar reason we exclude the object $\rm SDSS J0840+5624$ from our analysis of the proximity zones. In this spectrum we found an
absorption system with associated \lya and metal line absorption close to the quasar that attenuates enough flux such that it results in a spuriously small proximity zone. In this case, as in the case of BAL quasars, an absorption system associated with the quasar itself contaminates the proximity zone and thus we exclude this object from our analysis.
Note that we do not model such
intrinsic absorption line systems associated with the
quasar or its host galaxy  in the radiative transfer simulations (see \S~\ref{sec:sims}), such that any comparison would result in biases if we were not excluding these objects. 
We further discuss the spectrum of this excluded object
$\rm SDSS J0840+5624$ in Appendix~\ref{sec:J0840}.

\section{Methods}\label{sec:methods}

\subsection{Quasar Continuum Normalization}

\begin{figure*}[ht]
\centering
\includegraphics[width=\textwidth]{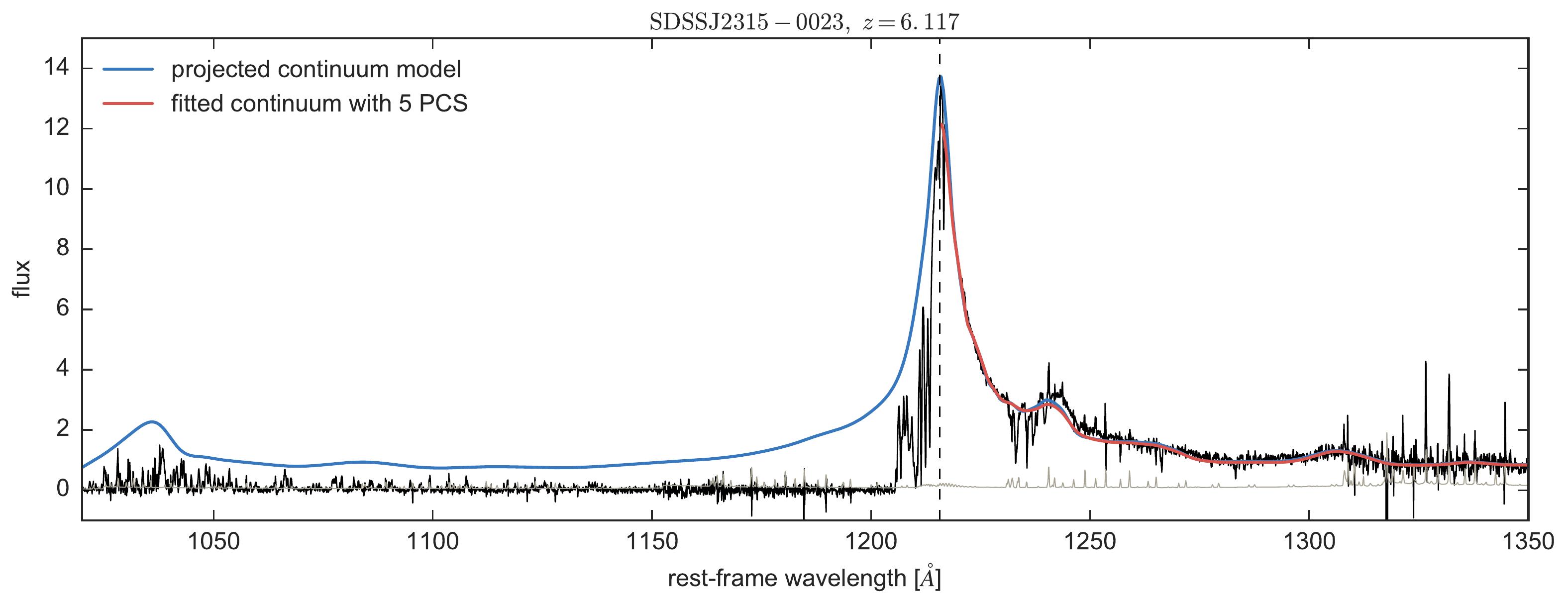}
\caption{Example of a quasar spectrum and its continuum model. The
  continuum model is fitted to the data with five PCS from \citet{Paris2011}
  on the red wavelength side of the spectrum (red curve),
  i.e. $1215.67$~{\AA}$\leq\lambda_{\rm rest}\leq 1600$~{\AA}, and then projected
  onto coefficients for a different set of PCS resulting in a continuum model also covering the
  blue wavelength side (blue curve), i.e. the whole spectral region between
  $1020$~{\AA}$\leq\lambda_{\rm rest}\leq 1600$~{\AA}, in order to predict the
  continuum level in the \lya forest region. \label{fig:continuum}}
\end{figure*}

We normalize all quasar spectra to unity in a region free of emission lines at $\lambda_{\rm rest}=1280$~{\AA}. Note that for a handful of quasars, that have very low signal-to-noise ratio data at this wavelength, we normalize
their spectra to unity at a slightly different wavelength. 
We then estimate the quasar continua with principal component spectra (PCS) from a principal component analysis (PCA) of low redshift quasar spectra \citep{Suzuki2006, Paris2011}. 
The idea of the PCA is to represent the continuum spectrum $\ket{q_{\lambda}}$ of a quasar by a reconstructed spectrum that consists of a mean spectrum $\ket{\mu_{\lambda}}$ and a sum of $m$ weighted PCS $\ket{\xi_{\lambda}}$, where the index $\lambda$ denotes the wavelength. Hence
\begin{align}
\ket{q_{\lambda}}\approx\ket{\mu_{\lambda}} + \sum_{i=1}^m \alpha_{i}\ket{\xi_{i, \lambda}},\label{eq:pca}
\end{align}
where the index $i$ refers to the $i$th PCS and $\alpha_{i}$ is its weight. For the majority of our continuum fits,
we use the PCS from \citet{Paris2011} that were derived to characterize quasar continua from $78$ high-quality spectra of bright $z_{\rm em}\sim3$ quasars in the Sloan Digital Sky Survey (SDSS).

The quasars in our sample are all at very high redshift and thus
experience absorption due to the intervening residual neutral hydrogen
along the line of sight bluewards of the \lya emission line, i.e. in
the \lya forest.  Thus we estimate the quasar continuum solely from
wavelengths redwards of rest-frame \lya which suffer only from modest
absorption by metal lines. \citet{Paris2011} provide a set of PCS for
wavelengths $1215.67$~{\AA}$\leq\lambda_{\rm rest}\leq 1600$~{\AA},
whose coefficients $\bm{\alpha_{\rm red}}$ we estimate by $\chi^2$ minimization using the noise vector from the spectra.

We use a projection matrix $\bm{P}$ to transfer the estimated coefficients for the PCS \textit{redwards} of Ly$\alpha$, $\bm{\alpha_{\rm red}}$, onto coefficients $\bm{\alpha}$ for a set of PCS that cover the \textit{entire} spectral region between $1020$~{\AA}~$\leq\lambda_{\rm rest}\leq 1600$~{\AA}, i.e. 
\begin{align}
\bm{\alpha} = \bm{P} \cdot \bm{\alpha_{\rm red}}.  
\end{align}
This projection matrix $\bm{P}$ has been calibrated by \citet{Paris2011} using the set of PCS for both the red wavelength side only and the whole spectral region covering wavelengths bluewards and redwards of \lya. 

Eqn.~(\ref{eq:pca}) gives us a model for the continuum of each quasar that includes the (absorbed) blue side of the spectrum. An example of a quasar spectrum from our data set and its continuum model is shown in Fig.~\ref{fig:continuum}. Note that in most cases we take five PCS on the red side into account to estimate the continuum model. However, a visual inspection of the continuum models
reveals that for four quasars a continuum model with three or seven PCS results in a better fit. 

Our PCA modeling of quasar continua is limited by the finite number of
principal components and the selection of quasars that were used to
create the PCS. For a few of the quasars in our sample, the PCS from
\citet{Paris2011} do not provide an acceptable
fit to the continuum, particularly for the fainter quasars in our sample. For these objects we instead use the set of PCS provided by \citet{Suzuki2006}, based on fainter lower redshift quasars,  
which qualitatively provide better fits to this handful of objects. 
When using the \citet{Suzuki2006} PCS basis, we determine the continuum
in the Ly$\alpha$ forest region using the method they advocate. Namely,
we determine the continua
at blue wavelengths $\lambda_{\rm rest} < 1216.67\,{\rm \AA}$ using the
best-fit continua to the red-side  $\lambda_{\rm rest} > 1216.67\,{\rm \AA}$.

The details of the continuum normalization are listed in Table~\ref{tab:continuum} in Appendix~\ref{sec:details_cont}.

\subsection{Measuring the Sizes of Quasar Proximity Zones}\label{sec:measure_prox_zones}

\begin{figure*}[h!]
\centering
\includegraphics[width=\textwidth]{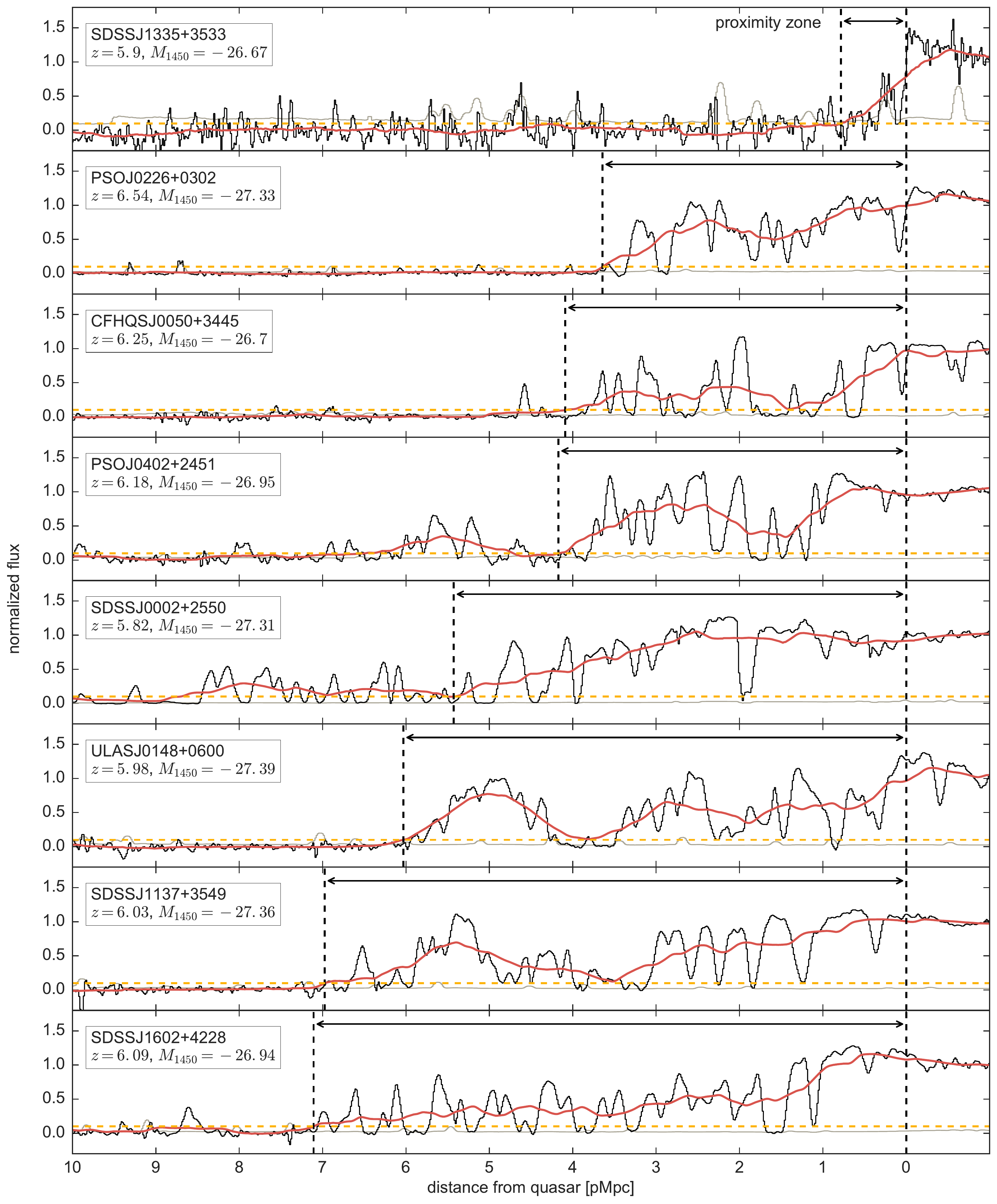}
\caption{Continuum normalized spectra of a subset of bright quasars in our data set with luminosities between $-27.5 \leq M_{\rm 1450} \leq -26.5$ showing the transmission profile within $10$~pMpc from the quasar. A boxcar smoothing of two pixels has been applied to the spectra (black) and their noise vectors (gray). 
The red curves show the quasar spectra smoothed to a resolution of $20${\AA} in observed wavelength. The horizontal yellow dashed lines indicate a flux level of $10\%$. 
The vertical black dashed lines show the extent of the quasar proximity zone from the quasar redshift (right) to the first drop of the smoothed flux below the $10\%$ level (left). \label{fig:9spectra}} 
\end{figure*}

\begin{figure*}[h!]
\centering
\includegraphics[width=\textwidth]{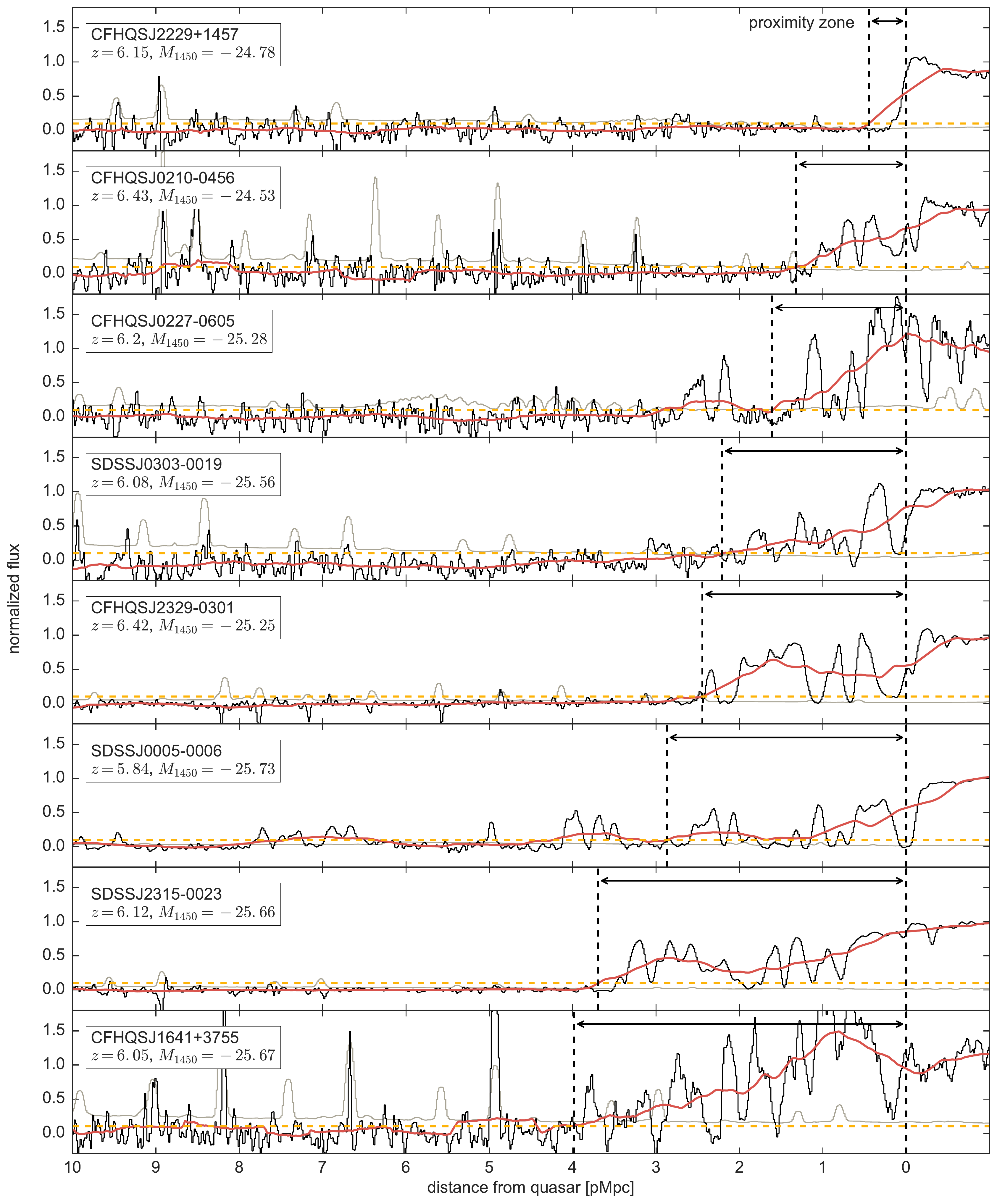}
\caption{Same as Fig.~\ref{fig:9spectra}, for a subset of faint quasars in our data set with luminosities between $-25.8 \leq M_{\rm 1450} \leq -24.5$. 
\label{fig:9spectra_faint}} 
\end{figure*}

In order to calculate the proximity zone sizes for each quasar we
adopt the standard definition used in the
literature \citep[see][]{{Fan2006}, {Willott2007}, {Willott2010}, {Carilli2010}, {Venemans2015}}.
Namely, we take the continuum normalized quasar spectra and smooth them with a
$20${\AA}-wide (observed frame) boxcar function. This smoothing scale corresponds to $\approx 1.0$~pMpc or a $\approx 705$~km/s window at $z=6$. 
We define the proximity zone
size as the distance to the first of three
consecutive pixels\footnote{We checked whether it
  changes the measurements of the proximity zone sizes when requiring up to
  ten pixels to be below the $10\%$ level, but we do not observe a
  significant difference.}
on the blue
side of the \lya emission line that show a 
drop of the smoothed flux below the $10\%$
level. This is demonstrated in Fig.~\ref{fig:9spectra} and
Fig.~\ref{fig:9spectra_faint} for subsets of bright ($-27.5 \leq M_{\rm 1450} \leq -26.5$) and faint ($-25.8 \leq M_{\rm 1450} \leq -24.5$) quasars
from our data sample, respectively. The depicted quasars all cover a similar range in luminosity, but nevertheless show a wide range of proximity zone sizes, between $0.8$~pMpc~$\lesssim R_p\lesssim 7.1$~pMpc for bright quasars and $0.5$~pMpc~$\lesssim R_p\lesssim 4.0$~pMpc for fainter quasars. Similar plots for the remaining objects in our quasar sample are shown in Appendix~\ref{sec:remaining_zones}. 
All measurements of the proximity zone sizes $R_p$ for our data set are listed in Table~\ref{tab:overview}.

\begin{figure}[th]
\centering
\includegraphics[width=.5\textwidth]{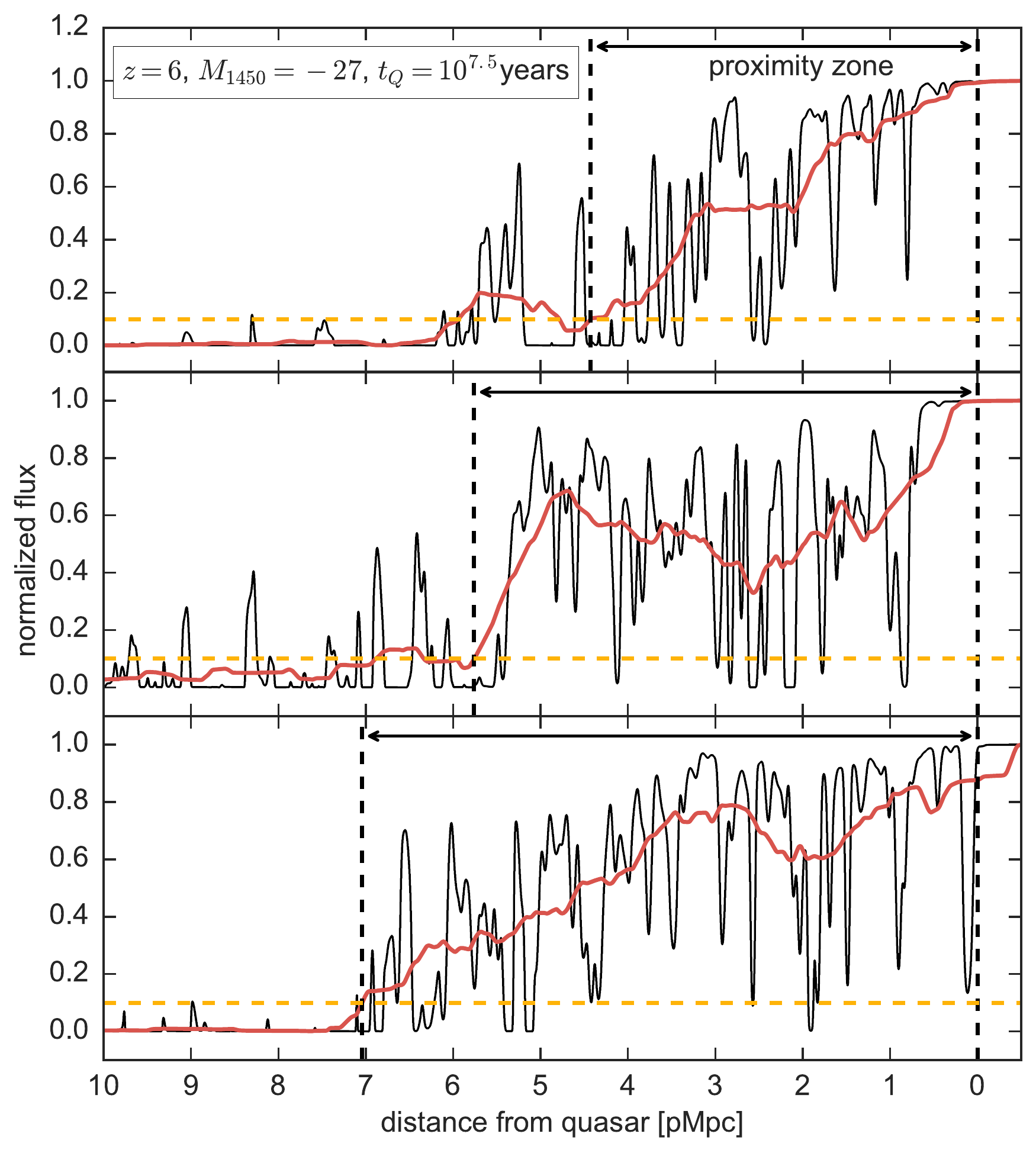}
\caption{Mock spectra from three different skewers through the radiative transfer simulation box for a quasar at $z=6$, a magnitude of $M_{\rm 1450}=-27$, and an age of $t_{\rm Q}=10^{7.5}$~yr, in a highly ionized IGM. The red curve shows the smoothed spectra and the black dashed lines indicate the extent of the proximity zones. \label{fig:mock_spectra}} 
\end{figure}

\section{Radiative Transfer Simulations}\label{sec:sims}

To interpret our measurements of $R_p$, we run a series of 
radiative transfer simulations of the effect of quasar ionizing photons on the IGM along
the line of sight similar to those performed by \citet{BoltonHaehnelt2007a}.  We apply the one-dimensional ionizing radiative transfer 
code from \citet{Davies2016} to skewers from a $100$~Mpc$/h$ Eulerian
hydrodynamical simulation run with the Nyx code \citep{Almgren2013,
  Lukic2015} at $z=6$ with $4096^3$ baryonic (Eulerian) grid elements and dark
matter particles. 
The radiative transfer computes the time-dependent ionization and
recombination of six species ($e^-$, \ion{H}{1}, \ion{H}{2}, \ion{He}{1},
\ion{He}{2}, \ion{He}{3}) as well as the associated photoionization
heating and cooling by various processes including adiabatic cooling
due to the expansion of the Universe and inverse Compton cooling off
cosmic microwave background (CMB) photons \citep[see][for
  details]{Davies2016}. We use $900$ skewers of density, temperature, and peculiar
velocity drawn along the $x$, $y$, and $z$ grid axes from the centers
of the $150$ most massive dark matter halos in the Nyx simulation,
corresponding to halo masses $M_{\rm h} \ga 4\times10^{11}$
M$_\odot$\footnote{In detail we find that starting the skewers from
  more or less massive halos has a negligible impact on the
  resulting proximity zone sizes, in agreement with \citet{BoltonHaehnelt2007a} and \citet{Keating2015}.}. 
Given the relatively coarse resolution of the Nyx simulation ($25$ kpc$/h$) and the lack of metal-line cooling, star formation, and feedback processes, we do not expect to resolve or accurately model dense gas inside of galaxies. For this reason we have removed one object in our sample ($\rm SDSS J0840+5624$; see Appendix~\ref{sec:J0840}) which exhibits strong associated metal line absorption at the quasar redshift that likely arises from dense gas in the vicinity of the quasar or from its host halo.

Quasar absolute magnitudes were converted to specific luminosity at
the hydrogen ionizing edge ($\nu = \nu_{\rm HI}$) using the
\citet{Lusso2015} spectral template, and for consistency with previous
studies we assume that the spectrum at $\nu > \nu_{\rm HI}$ behaves as
a power law\footnote{Note that the magnitudes for the quasars in our data sample have been calibrated by \citet{Banados2016} 
 according to the template from \citet{Selsing2016} instead of \citet{Lusso2015}. However, due to the similarities between the two templates at observed wavelengths (redward of \lya), we expect only a very minor inconsistency in the inferred ionizing luminosity.} with $L_\nu \propto \nu^{-1.7}$. 
We assume that the quasar
turns on abruptly 
and emits at constant luminosity for its
entire age $t_{\rm Q}$, i.e. a so-called ``light bulb" model. For the quasar 
age
we assume a fiducial value of $t_{\rm Q}= 10^{7.5}$~yr, but we later investigate
the dependence of $R_p$ on $t_{\rm Q}$ in \S~\ref{sec:lifetime}.
We consider two initial conditions for the ionization state of gas in the
simulation: either the gas is initially highly ionized by a uniform ionizing background
 or the gas is initially completely neutral. In the highly ionized case, we add ionizing radiation due to the ultraviolet background (UVB) 
 to each cell leading to an ionization rate $\Gamma_{\rm UVB}=2\times 10^{-13}$ s$^{-1}$,
 consistent with observations of Ly$\alpha$ forest opacity at $z\sim6$ \citep[e.g.][]{WyitheBolton2011}
 and resulting in a neutral fraction of $f_{\rm HI}=1.5\times10^{-4}$, 
 with a spectrum characteristic of galactic sources (i.e. a sharp cutoff above 
the \ion{He}{2} ionizing edge). For simplicity we assume that $\Gamma_{\rm UVB}$ is
constant with redshift, but we note that the resulting proximity zone sizes are insensitive
to changes in $\Gamma_{\rm UVB}$ of a factor of a few, 
a non-trivial point that we will discuss in more detail in future work \citep{Daviesinprep}. 
We compute radiative transfer outputs for quasars of varying
luminosity and redshift 
bracketing the properties of our observed quasar sample in 
in both ionized and neutral scenarios. We assume here that the
  \textit{overdensity} field does not significantly evolve with
  redshift from our $z=6$ output across the redshift range we study,
  and simply re-scale physical densities by $(1+z)^3$.

Ly$\alpha$ forest spectra are computed by combining the neutral fraction $f_{\rm HI}$ and
gas temperatures 
from the radiative transfer simulation with the
velocity field from the Nyx simulation, summing the absorption from
each gas element using the efficient Voigt profile approximation of
\citet{TepperGarcia2006}. 
Finally, to retrieve $R_p$ from the
Ly$\alpha$ forest spectra, we perform a similar analysis as is applied
to the real spectroscopic data: we smooth the spectra with a boxcar
filter of $20$~{\AA} in the observed frame, and then locate where the
transmitted flux first drops below $10\%$. 
A few skewers through the simulation box giving mock \lya forest spectra in a highly ionized IGM are shown in Fig.~\ref{fig:mock_spectra}. 

In this work we present the general trends of $R_p$ with quasar parameters, leaving a more detailed discussion of the structure and evolution of these simulated proximity effect spectra to future work \citep{Daviesinprep}.

\section{The Redshift Evolution of Proximity Zone Sizes}\label{sec:z_evolution}

In this section we will present our measurements of the proximity zone
sizes for the ensemble of quasar spectra in our data set. Since these
quasars cover a wide range of luminosities, we first put them on a
common luminosity scale, in order to facilitate
a study of the redshift evolution of proximity zone sizes. In the last part of this section we
compare our measurements to previous work.

\subsection{Correcting for the Quasar Luminosity}

We expect the size of the proximity zone of a quasar to depend on
their luminosity. Proximity
zone sizes may also evolve with redshift in a way that tracks the
evolution of the ionization state of the IGM driven by the underlying
evolution in the UVB. To study the evolution of the proximity zone size with redshift, we have to remove
the dependency on quasar luminosity, and normalize our measurements to the same
absolute magnitude. The approach that has been adopted in the
literature, is to re-scale all quasar luminosities to a common value
assuming a particular model for the luminosity dependence, in order to
remove the large scatter in proximity zone sizes driven by the wide
range of quasar luminosities present in the data set. This then
enables one to study the redshift evolution of proximity zone sizes. However, the scaling of the proximity zone size with luminosity depends on the physical conditions of the ambient IGM. 
In this section we study radiative transfer simulations to better understand the expected
luminosity scaling, and its dependence on the IGM ionization state. We
compute the luminosity scaling in our data set and compare to these simulations.
We use these simulations to advocate for an approach to luminosity correct our measurements
and search for a redshift evolution.  

In general, the luminosity dependence of proximity zone sizes could
depend on the ionization state of the surrounding IGM.  Specifically
eqn.~\ref{eq:rp} indicates that the size of an ionization front $R_{\rm
  ion}$ expanding into a neutral IGM scales as $R\propto
\dot{N}_{\gamma}^{1/3}$. On the other hand, using an analytical model
\citet{BoltonHaehnelt2007a} showed that the proximity zone size
scales as $R_{p}\propto \dot{N}_{\gamma}^{1/2}$, if the IGM is in fact highly ionized by
either the quasar itself or the UVB. To better understand this
scaling, we turn to our radiative transfer simulations (see
\S~\ref{sec:sims}), which were run for an ensemble of quasars that
span the same range of luminosities as our data set. 
Note that the redshifts of the simulated quasars and the IGM are set to $z=6$.

\begin{figure}[t]
\centering
\includegraphics[width=.5\textwidth]{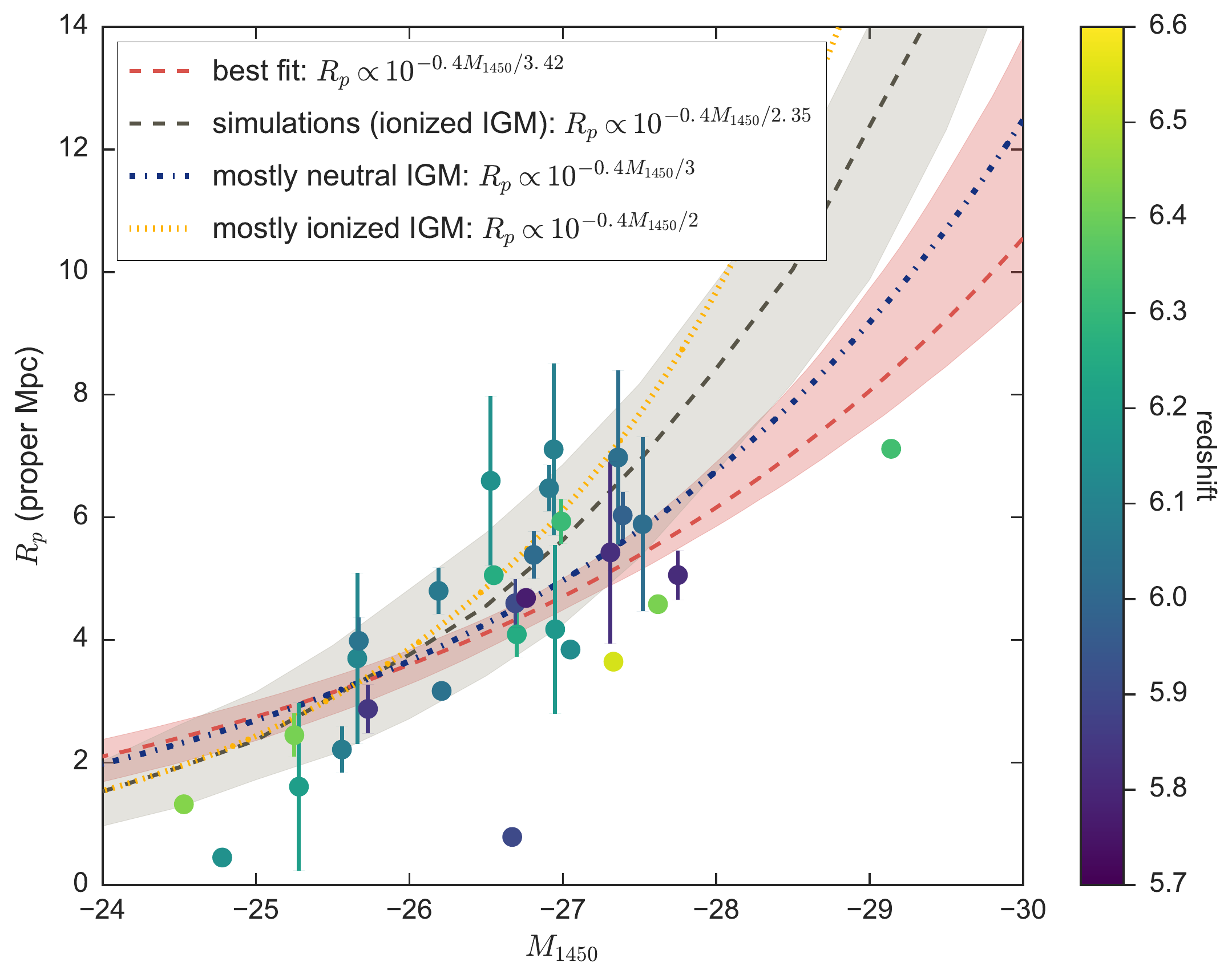}
\caption{Sizes of the proximity zones shown dependent on the quasars' magnitude $M_{\rm 1450}$, color coded by their redshifts. The red dashed line shows the best power-law fit to the measurements with a $1\sigma$-uncertainty level from bootstrapping errors. The gray dashed line shows the expected evolution of the proximity zones from radiative transfer simulations in a highly ionized IGM. The blue dashed-dotted and yellow dotted curves show the theoretical expectations when the IGM surrounding the quasars is neutral or highly ionized, respectively. \label{fig:RpvsM}} 
\end{figure}

The simulated relation between proximity zone size and quasar magnitude for a highly ionized IGM is shown as the gray dashed line in Fig.~\ref{fig:RpvsM}, with
the shaded region illustrating the $1\sigma$ scatter of simulated sizes due to cosmic
variance alone (i.e. no measurement error). 
The relation from the simulations in a highly ionized IGM is reasonably well fit
by the power-law
\begin{align}
R_p \approx 5.57\,{\rm pMpc}\times 10^{-0.4M_{\rm 1450}/2.35}. \label{eq:R_pvsM}
\end{align}

Our simulations suggest that the proximity zone sizes in a highly
ionized IGM scale as $R_p\propto \dot{N}_{\gamma}^{1/2.35}$, which
lies in between the two theoretically expected relations for a neutral
or mostly ionized ambient IGM, shown as the blue dashed-dotted and yellow
dotted curves, respectively. 
Note that the normalization for these two analytic
curves has been arbitrarily chosen such that all curves intersect at
the same point. The simulated relation does not align exactly with the analytically
expected relation in a highly ionized IGM, possibly due to heating effects of
\ion{He}{2} reionization.

Note however, that we obtain a very similar scaling from our radiative transfer simulations, when assuming a 
neutral ambient IGM:
\begin{align}
R_p \approx 5.03\,{\rm pMpc}\times 10^{-0.4M_{\rm 1450}/2.45}. 
\end{align}
This is due to the fact that the measured proximity zone sizes $R_p$ do not trace the extent of the ionization front $R_{\rm ion}$ described in eqn.~(\ref{eq:rp}) for an assumed quasar age of $t_{\rm Q}\sim10^{7.5}$~yr, but end much earlier within the ionized \hii region around the quasar. Because the IGM is indeed
highly ionized in this region, one expects a scaling 
similar
to that of a highly ionized IGM. Hence, it is no surprise to find the two relations from the radiative transfer simulations in a highly ionized ambient IGM as well as in an originally neutral ambient IGM to behave similarly. 
We will come back to this in more detail in \S~\ref{sec:shallow_evolution}.

Given this new found intuition from radiative transfer simulations, it is now interesting
to study the luminosity scaling of proximity zone sizes in our data. 
To this end, we show the dependence of our measured proximity zone sizes
on the quasar's magnitude, which is proportional to the quasar luminosity, in Fig.~\ref{fig:RpvsM}. Measurements have been
color coded by the emission redshift of the quasar. 
The errorbars of the measurements reflect the uncertainty on the proximity zone
sizes due to the uncertainty of the quasar redshift (see \S~\ref{sec:data}). 
We fit a power-law to our measurements and obtain 
\begin{align}
 R_p\approx 4.71\,{\rm pMpc}\times 10^{-0.4\times (M_{\rm 1450}+27)/3.42}, \label{eq:bestfit} 
\end{align}
as the best fit, shown as the red dashed line. The
$1\sigma$ uncertainty of this fit is given by the shaded region, which was
determined by bootstrap re-sampling of our measurements with replacement, and repeating the fit
$1000$ times. Note that we do not weight our measurements by the redshift errors
in this fit.
In general, we find reasonable agreement between the scaling relations from the radiative transfer simulations and our data.
The data seem to favor a shallower evolution, but this is complicated by the fact
that we fit a sample over a large range of redshifts, and also, possibly, that
exceptionally small proximity zones (see \S~\ref{sec:small_zones}) are pulling down the fit.

So which relation should we adopt to put the measured proximity zone sizes onto a
common luminosity scale?  Due to the large body of evidence for the
IGM being mostly ionized at $z\sim 6$ \citep[e.g.][]{WyitheBolton2011,
  Calverley2011, Becker2015, McGreer2015} we choose the 
relation from our radiative transfer simulations assuming a highly ionized IGM
(eqn.~(\ref{eq:R_pvsM})) to eliminate the dependence on quasar luminosity and
normalize our
measurements to a common magnitude of $M_{\rm 1450}=-27$. However, our
results are only marginally dependent on this choice, since the luminosity
scaling for an ionized and neutral IGM 
are in fact very similar, and the main conclusions of this paper are not influenced
by this choice. Thus we re-scale our measured
proximity zone sizes with 
\begin{align}
R_{p, \rm corr} \approx R_p\cdot 10^{-0.4(-27-M_{\rm 1450})/2.35}. \label{eq:correction} 
\end{align}

Our measurements of the quasar proximity zone sizes are provided
in Table~\ref{tab:overview}, with the two rightmost columns showing the measured
proximity zone $R_p$ in proper Mpc and the corrected proximity zone size
$R_{p, \rm corr}$, which are re-scaled to an absolute
magnitude of $M_{\rm 1450}=-27$ following eqn.~(\ref{eq:correction}).

\begin{figure*}[t]
\centering
\includegraphics[width=\textwidth]{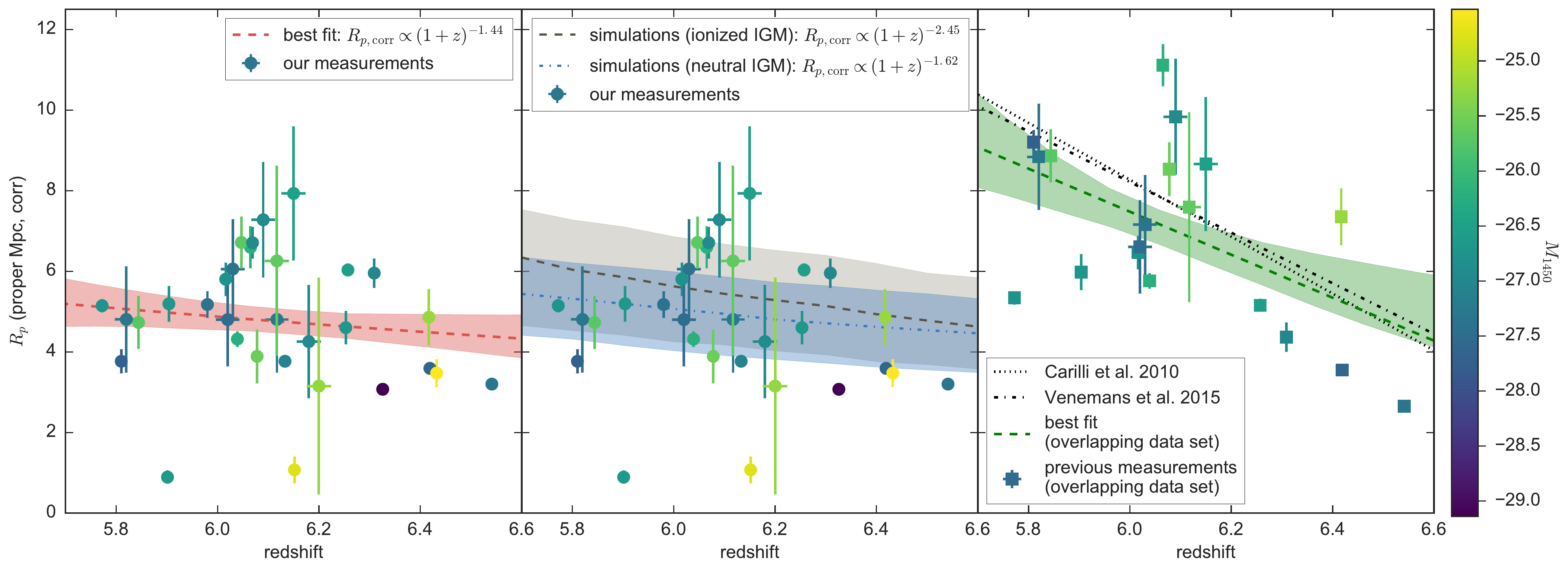}
\caption{Evolution of the luminosity corrected proximity zone sizes with redshift, color coded by the quasars' actual magnitude $M_{1450}$. The data points showing our measurements for $R_{p,\rm corr}$ in both the right and middle panel are the same. In the left panel, the red dashed line shows the best power-law fit to the measurements with a $1\sigma$-uncertainty level determined by bootstrapping. In the middle panel the gray dashed line shows the evolution of the proximity zones found in radiative transfer simulations when assuming a highly ionized IGM, whereas the blue dashed line is the result of the radiative transfer simulations assuming a mostly neutral ambient IGM. The shaded areas show the respective $1\sigma$ uncertainties of the relations due to cosmic variance. In the right panel the square data points show previous measurements by \citet{Carilli2010} and \citet{Venemans2015} of the quasars in common between our and their data sets, after correcting with updated redshift and magnitude estimates. The green dashed line shows the best fit to these measurements with a $1\sigma$-uncertainty level determined by bootstrapping. The black dotted and dash-dotted curves show linear fits to the measurements of the whole data set from \citet{Carilli2010} and \citet{Venemans2015}, respectively. \label{fig:Rpvsz}} 
\end{figure*}

\subsection{The Redshift Evolution of Quasar Proximity Zone Sizes}

After correcting our measurements for the quasar luminosity, we can study the evolution of the proximity zones with redshift. 
In Fig.~\ref{fig:Rpvsz} we show this redshift evolution of our luminosity corrected measurements, color coded by $M_{\rm 1450}$, in both the left and middle panel. 
In the left panel, the red dashed line shows the best-fit power-law fit
to the redshift evolution
\begin{align}
R_{p, \rm corr} \approx 4.87\,{\rm pMpc} \times\left(\frac{1+z}{7}\right)^{-1.44}. 
\end{align}
The shaded region indicates the $1\sigma$ uncertainty on the fit
determined by bootstrap re-sampling of our measurements
with replacement and repeating the fit $1000$ times.

In the next section we use our radiative transfer simulations to better understand
the redshift evolution of proximity zones, and
show that the shallow redshift evolution that we measure is indeed 
expected.  However,
the black dashed and dash-dotted lines in the right panel of Fig.~\ref{fig:Rpvsz} are linear fits to the measurements of quasar proximity zones from
previous analyses of similar data sets by \citet{Carilli2010} and \citet{Venemans2015}, which show a much steeper trend. We compare
our measurement to previous work in \S~\ref{sec:comparison}, and discuss the sources
of this discrepancy.

\subsection{Understanding the Shallow Redshift Evolution of Quasar Proximity Zone Sizes}\label{sec:shallow_evolution}

We use our radiative
transfer simulations to investigate the expected redshift evolution of
proximity zone sizes for both a highly ionized ambient IGM and a neutral IGM. Our results are shown in the middle panel of Fig.~\ref{fig:Rpvsz}
as the gray dashed line (ionized) and blue dash-dotted line (neutral),
respectively. 
In the case of an ionized ambient IGM the corrected quasar proximity zones follow a power-law
\begin{align}
R_{p, \rm corr} \approx 5.66\,{\rm pMpc}\times \left(\frac{1+z}{7}\right)^{-2.45},  
\end{align}
whereas the evolution in the case of a neutral IGM is well fit by 
\begin{align}
R_{p, \rm corr} \approx 5.06\,{\rm pMpc}\times \left(\frac{1+z}{7}\right)^{-1.62}. 
\end{align}
The shaded regions around the curves show the scatter about
these relations due to cosmic variance
for quasars with magnitude $M_{\rm 1450}=-27$, but they slightly over- and underpredict the scatter for higher and lower luminosity quasars, respectively. 
The two scenarios of an ionized and a neutral IGM both result in a relatively
shallow redshift evolution of proximity zones and are thus consistent with our
measured redshift evolution shown in the left panel as the red dashed curve, but inconsistent with the steep
evolution found in previous work (right panel).

Why should the redshift evolution of quasar proximity zones be so
shallow?  The end of the proximity zone around quasars is defined as
the location where the smoothed transmitted flux drops below the
$10\%$ transmission level and a corresponding limiting optical depth of $\tau_{\rm lim}=2.3$ is reached.
\citet{BoltonHaehnelt2007a} determine in eqn.~(8) of their paper the ionization rate $\Gamma_{\rm lim}$ that is necessary to produce a neutral gas fraction $f_{\rm lim}$ that results in the required limiting optical depth $\tau_{\rm lim}$. Assuming a highly ionized gas in ionization equilibrium and making reasonable assumptions about the gas temperature of the IGM at $z\sim 6$, one obtains an ionization rate of $\Gamma_{\rm lim} \sim 4\times 10^{12}$~s$^{-1}$ at the end of the proximity zone $R_p$, which is an order of magnitude larger than ionization rate of the UVB $\Gamma_{\rm UVB}\sim 2\times 10^{-13}$~s$^{-1}$ at this redshift \citep{WyitheBolton2011}. Thus the total ionization rate at the end of the proximity zone $\Gamma_{\rm lim}$, which is the sum of the ionization rate of the background radiation $\Gamma_{\rm UVB}$ and the ionization rate of the quasar itself $\Gamma_{\rm QSO}$, has to be totally dominated by the latter.

As such, in our simulations $R_p$ is essentially independent of
$\Gamma_{\rm UVB}$ and hence fairly insensitive to the neutral
fraction $f_{\rm HI}$ of a highly ionized IGM, provided
the quasar has been emitting light for longer than the equilibration timescale $t_{\rm eq}$
of the gas, 
which denotes the timescale on which the gas reaches ionization
equilibrium (see \S~\ref{sec:lifetime}). The quasar age thus has to be
$t_{\rm Q} \gtrsim t_{\rm eq} \sim 1/\Gamma_{\rm QSO}(R_p) \sim
10^{5}$~yr, in order for the ambient gas to have reached ionization
equilibrium.

If the IGM surrounding the quasars is instead very neutral, i.e. $f_{\rm HI}\sim 0.1-1.0$, eqn.~(\ref{eq:rp}) for the location of the ionization front
indicates that the size of the ionized \hii region around the quasar scales as $R_{\rm ion}\propto (t_{\rm Q}/f_{\rm HI})^{1/3}$. This suggests that the proximity zones could be sensitive to the neutral gas fraction (subject to a degeneracy
with the quasar age). 
Indeed for short ages the measured proximity zone $R_p$ will trace the expanding ionization front $R_{\rm ion}$ around the quasar, i.e. $R_p\approx R_{\rm ion}$, which increases with the quasar age. However, $R_p$ will cease to grow
further once it reaches a distance given by the $10\%$ transmission level of
the flux, according to the definition of $R_p$. Thus even as $R_{\rm ion}$
continues to grow with age, the proximity zone size saturates, and will
be insensitive to both the quasar age and neutral fraction. 
This implies that the measured proximity zone size provides a lower limit
on the location of the ionization front $R_{\rm ion}$. 
The maximum size of the proximity zone $R_{p, \rm max}$ is given by the distance at which a $10\%$ flux transmission level corresponding to a limiting optical depth of $\tau_{\rm lim}=2.3$ is reached. 
The age of the quasar at this distance is $t_{\rm Q} > t_{\rm Q} (R_{p, \rm max}) \sim 10^6$~yr under the assumption of a homogeneous IGM \citep[see eqn.~(13) in ][]{BoltonHaehnelt2007a}.

Thus for a highly plausible quasar age of $t_{\rm Q}\sim 10^{7.5}$~yr
\citep[see e.g.][]{Martini2004}
the proximity zones $R_p$ in a neutral IGM will look the same as for the highly ionized case, because the proximity zones by definition cease to grow once a flux transmission level of $10\%$ is reached, as has been
previously pointed out by \citep{BoltonHaehnelt2007a}.  This demonstrates the insensitivity of $R_p$ to the neutral fraction $f_{\rm HI}$ of the IGM for an appropriate choice of quasar age \citep[see also][for an analogous argument in the context of \ion{He}{2} proximity zones]{Khrykin2016}.

Regarding the redshift evolution of the proximity zone sizes, eqn.~(11) in \citet{BoltonHaehnelt2007a} predicts a very shallow scaling of $R_p\propto (1+z)^{-2.25}$ due to the density evolution in the universe. The simulations in Fig.~\ref{fig:Rpvsz} reveal a similar scaling, i.e. $R_p\propto (1+z)^{-2.45}$ for a neutral ambient IGM and $R_p\propto (1+z)^{-1.62}$ for a highly ionized IGM with a quasar age of $t_{\rm Q}\sim 10^{7.5}$~yr. The true scaling does not match exactly the analytically expected scaling due to  different heating effects.

Hence we conclude that the redshift evolution of $R_p$ is not a very useful probe of the ionization state of the IGM. In a highly ionized IGM due to the definition of proximity zones, the ionization rate at the end of the proximity zone is totally dominated by the ionization rate of the quasar itself and thus essentially independent of the background radiation and the ionization state of the surrounding IGM. In a neutral IGM, $R_p$ ceases to grow once a maximum size $R_{p, \rm max}$ corresponding to a limiting optical depth $\tau_{\rm lim}$ at the $10\%$ flux transmission level is reached. 
The observed shallow redshift evolution is thus perfectly consistent with our models for both a highly ionized and a neutral IGM for a quasar age of $t_{\rm Q}>10^6$~yr, in particular for a fiducial quasar age of $t_{\rm Q}\sim10^{7.5}$~yr. 
None of our simulations reveal the steep evolution that was measured by previous analyses \citep{Carilli2010, Venemans2015}. Our models do not predict any large proximity zones of $R_p\gtrsim 10$~pMpc that would be consistent with their best-fit curve at $z\approx 5.8$. We will now investigate possible reasons leading to the discrepancy between ours and previous analyses.

\subsection{Comparison to Previous Analyses}\label{sec:comparison}

\begin{figure}[t]
\centering
\includegraphics[width=.5\textwidth]{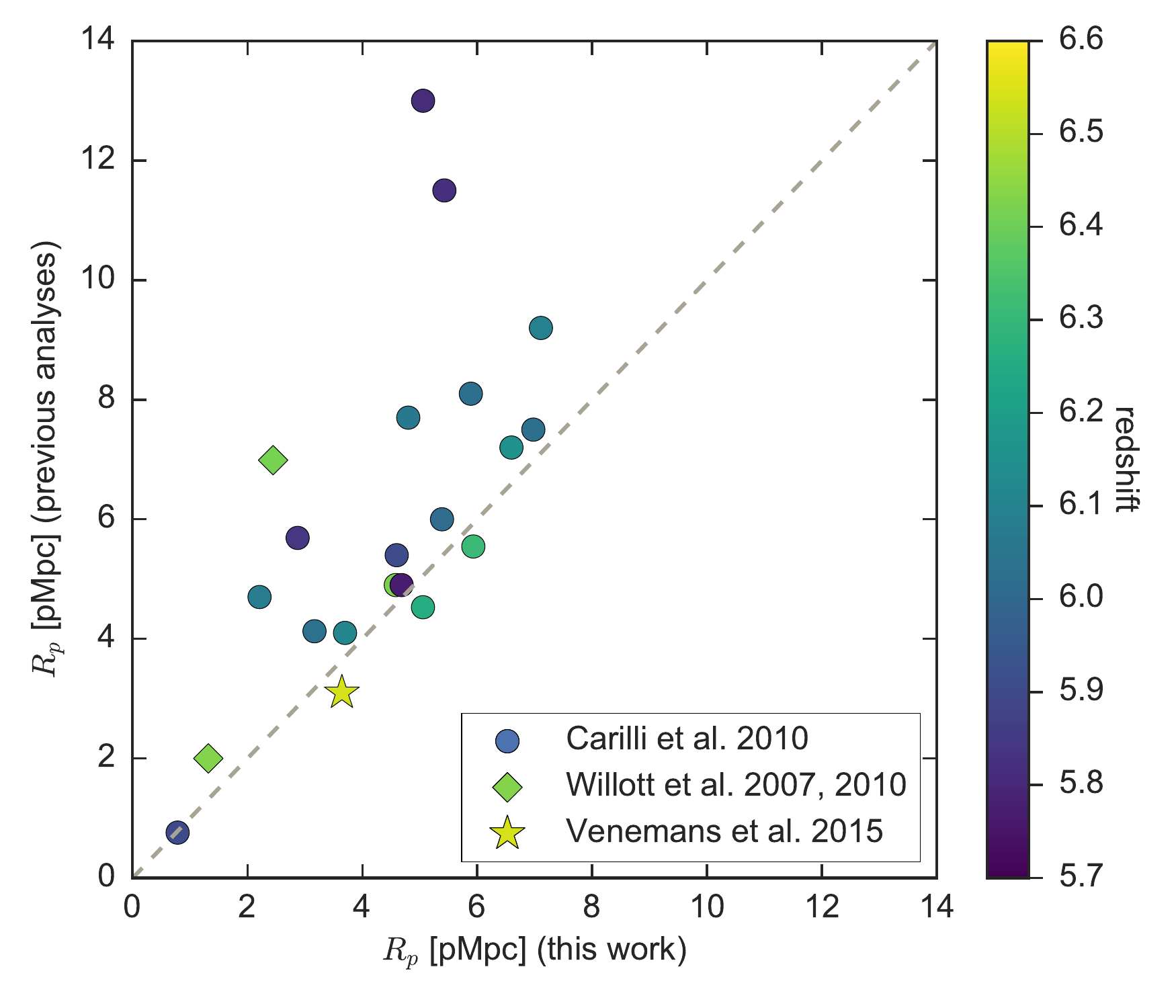}
\caption{Comparison of previous measurements of quasar proximity zones $R_p$ to the analysis presented in this paper for the overlapping objects in the various data sets. The gray dashed line shows the exact one-to-one relation. In general we recover smaller proximity zones than previous analyses, but mostly consistent within $\lesssim 2$~pMpc. \label{fig:carilli}} 
\end{figure}

\begin{figure*}[t]
\centering
\includegraphics[width=.9\textwidth]{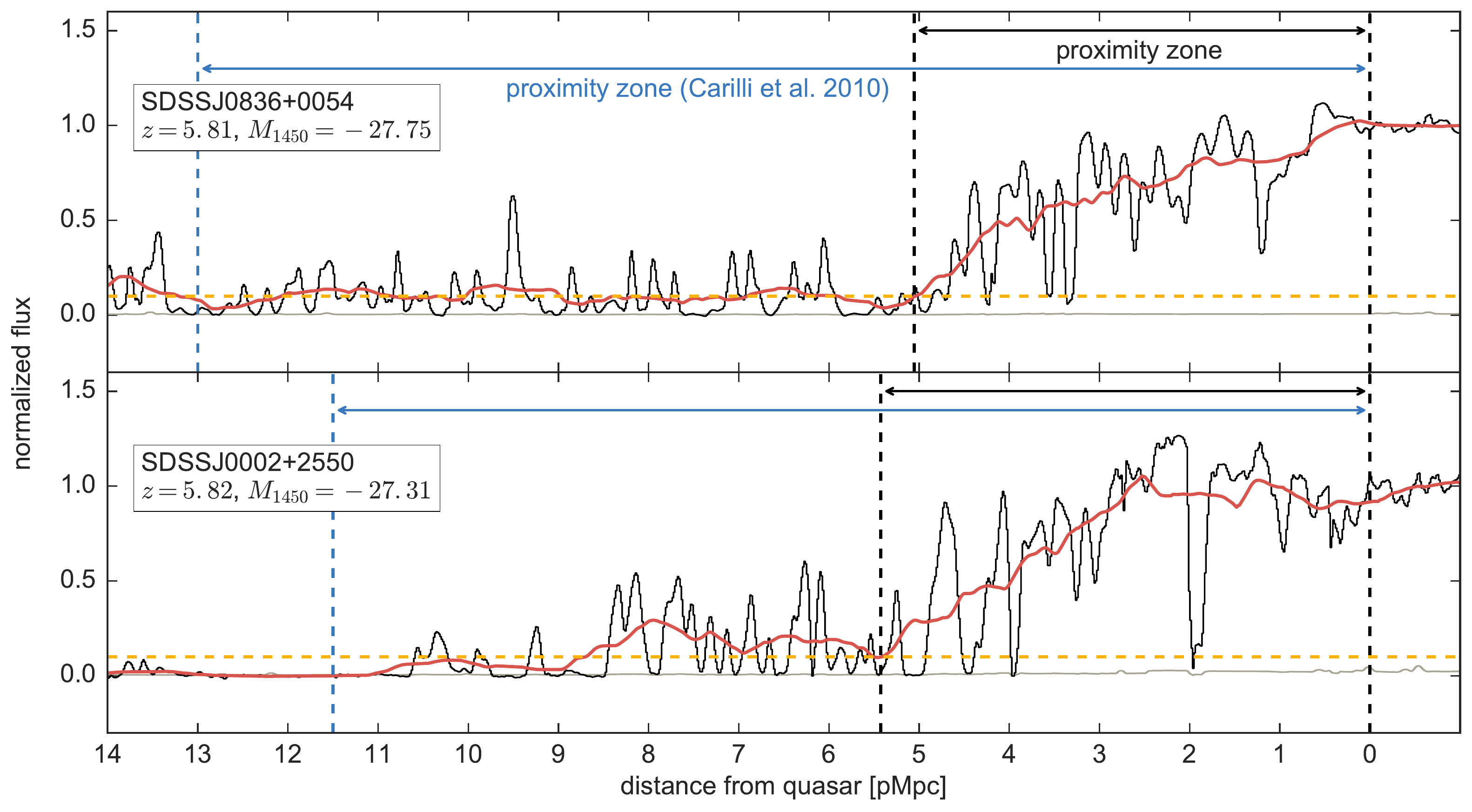}
\caption{Quasar spectra from our data set with the largest discrepancy between our measurements of the proximity zones and the measurements by \citet{Carilli2010}. \label{fig:large_zones}} 
\end{figure*}

There are several possible reasons causing the differences between
ours and previous analyses that result in the different measured
redshift evolution of quasar proximity zones: we use a different
scaling with luminosity applied to correct all proximity zone
measurements and put them on the same luminosity scale, different
quasar spectra\footnote{Although many of the objects are overlapping between ours and previous data sets, we co-added different exposures from different observing runs and have thus gained higher quality spectra. }, different approaches to fitting continua and measuring
the proximity zone sizes, and our measurements also rely on
updated luminosities and redshift measurements. Below
we discuss each of these differences and their impact on our results in turn.

To address the issue of different luminosity scalings we facilitate a
comparison to previous work by \citet{Carilli2010} and
\citet{Venemans2015} by taking their measurements of $R_p$ and
correcting them with our luminosity correction ($R_{p,\rm corr} \propto \dot{N}_{\gamma}^{1/2.35}$ instead of their $R_{p,\rm corr} \propto \dot{N}_{\gamma}^{1/3}$; see eqn.~(\ref{eq:correction}))
using updated measurements of $M_{\rm
  1450}$ from \citet{Banados2016} for their objects. In their work
they chose a linear relation 
between proximity zone size and
redshift, i.e. $R_p\propto z$, and thus we also show a linear fit to
their corrected measurements in the right panel of
Fig.~\ref{fig:Rpvsz} in order to simplify the comparison. Our best fit slope to the measurements of 
\citet{Carilli2010} is $m\approx -7.07$ (black dotted line) and $m\approx =-6.28$ (black dash-dotted line) for measurements by
\citet{Venemans2015}, which is considerably steeper than the slope
$m\approx-1.12$ that we obtain when applying a linear fit to our
data.
Thus we conclude that the difference in luminosity scaling does not cause the
discrepancy between the different analyses.

However, both previous analyses contain objects in their data sample that are not included in our sample. 
Although the data sets from \citep{Carilli2010} and \citet{Venemans2015} partially
overlap with our sample, they also contain distinct objects.
In order to verify that differences in the slope of
the redshift evolution 
are not being driven by differences in the quasar samples,
we restrict the analysis now to measurements of the $18$ quasars that are
common to our sample and those of \citet{Carilli2010} and
\citet{Venemans2015}. For five of these quasars,
our analysis uses updated redshift measurements and thus we
correct the measurements of $R_p$ from previous work for the difference
between the old and the new redshift measurements in order to have consistent redshifts and afterwards use eqn.~(\ref{eq:correction}) to obtain $R_{p, \rm corr}$ from their measurements of $R_p$. Their corrected measurements of the objects overlapping in both data sets are shown in the right panel of Fig.~\ref{fig:Rpvsz} and the best linear fit
for the redshift evolution is shown as the green dashed curve. It still shows a relatively steep slope of 
$m\approx -5.33$ although slightly shallower than before.
Thus we conclude that differences in the data sets are not driving the discrepancy in the results.

Note however, that the differences in the resulting redshift evolution of $R_p$ appear to be driven by a handful of objects, where we measure significantly different proximity zone sizes than previous work. 
In Fig.~\ref{fig:carilli} we compare our measurements of the (uncorrected) $R_p$ to those
measured by \citet{Willott2007}, \citet{Willott2010},
\citet{Carilli2010} (who updated the measurements originally performed
by \citet{Fan2006}), and \citet{Venemans2015} for all objects that are
overlapping between our data sample and these previous analyses. We have adopted consistent redshifts for all objects here in order to facilitate a one-to-one comparison that is not driven by changes in redshift, i.e. for the objects for which we have updated redshift measurements, we correct the previously analyzed proximity zones for the difference.

The measurements which lie along the gray dashed line indicate
agreement between the different analyses. In general, we measure
smaller sizes of the proximity zones than previously obtained, but for
most objects the measurements agree within $\Delta R_p\lesssim 2$~pMpc.
Despite the
fact the fact that for a significant fraction of these overlapping 
quasars the analyzed data comes from the same instrument (Keck/ESI), small differences in the measurements can be attributed to different data reduction pipelines and in many cases we have co-added data from different runs to get higher $\rm S/N$ spectra and thus the final quasar spectra might differ in quality. 
However, there are a few outliers in Fig.~\ref{fig:carilli}, for which
our measurements differ significantly from previous work, i.e. $\Delta R_p > 2$~pMpc, which we discuss in further
detail.

We measure much smaller proximity zones for three quasars in
particular: $\rm SDSS J0836+0054$ ($R_{p, \rm Carilli} \approx
13.0$~pMpc vs. our measurement $R_p=5.06\pm0.40$~pMpc) at $z=5.810$, $\rm SDSS
J0002+2550$, ($R_{p, \rm Carilli} \approx 11.5$~pMpc vs. our measurement
$R_p=5.43\pm1.49$~pMpc) at $z=5.82$, and $\rm CFHQS J2329-0301$, ($R_{p, \rm Willott} \approx 7.0$~pMpc vs. our measurement
$R_p=2.45\pm0.35$~pMpc) at $z=6.417$. 

The spectra of the first two objects, $\rm SDSS J0836+0054$ and $\rm SDSS
J0002+2550$ and our measured proximity zones (black dashed lines) as well as the
measurements from \citet{Carilli2010} (blue dashed lines) are shown in
Fig.~\ref{fig:large_zones}. Both analyses use the same redshift measurements and thus no differences due to redshift errors should be causing the discrepancy we see
here. 
The spectrum of $\rm SDSS J0836+0054$ in the upper panel shows
transmitted flux throughout the spectrum with the smoothed flux
oscillating around the $10\%$ level. Our spectrum nevertheless shows a
significant drop below the $10\%$ level at $R_p\approx
5.06$~pMpc. This measured $R_p$ does not change significantly when fitting the continuum model with a different set of PCA components.

For $\rm SDSS J0002+2550$ (lower panel) the size of the proximity zone is more sensitive to the precise placement of the quasar continuum level. 
A slightly different continuum normalization (i.e. taking a different set of PCA components) would increase the measurement of the proximity zone from $R_p\approx 5.43$~pMpc to $R_p\approx 8.85$~pMpc, but still falls considerably short of
the $R_p=11.5$~pMpc previous measurement by \citet{Carilli2010}. 
Another reason for the discrepancy in the $R_p$ measurements could be the higher signal-to-noise data in our sample. However, it remains unclear whether the differences in the continuum normalization and the better quality data could cause the whole discrepancy of the two measurements.

The third object that has a significantly different proximity zone measurement is $\rm CFHQS J2329-0301$, whose proximity zone was determined by \citet{Willott2007} to be $R_p\approx 6.3$~pMpc (after updating their measurements with a new redshift measurement it is now $R_p\approx 7.0$~pMpc). However, most of this discrepancy between their measurement and ours can be contributed to the fact that for this particular object, \citet{Willott2007} decided after inspecting the spectrum by eye that they take the \textit{second} drop of the smoothed flux below $10\%$ level as the end of the proximity zone instead of the \textit{first} drop according to the standard definition of proximity zones \citep{Fan2006}. They determine the first drop in flux below the $10\%$ transmission level to be at $R_p\approx 3.7$~pMpc (with an updated redshift measurements this would be $R_p\approx 4.4$~pMpc), which would be much more consistent with our measurement of $R_p=2.45\pm0.35$~pMpc.

Another reason for the discrepancy in the redshift evolution of the proximity zone measurements is caused by the exclusion of the quasar $\rm SDSS J1335+3533$ at $z=5.9012$ from previous analyses. 
For this particular object, which has a very small proximity zone of $R_p=0.78\pm0.15$~pMpc, our measured size agrees with the one measured by \citet{Carilli2010}. However, this object is a weak emission line quasar and has been
somewhat arbitrarily excluded from previous analyses due to this fact \citep[see][]{Carilli2010, WyitheBolton2011}. 
The reason given in the aforementioned papers for is the ``fundamentally different nature" of such objects. 
However \citet{Diamond-Stanic2009} showed that weak emission line quasars do not show significantly different UV continuum slopes
apart from the emission lines and thus do not differ in their physical properties. The original reason for \citet{Fan2006} to exclude this object from their analysis was that the only redshift measurement they had at that time came from the onset of strong \lya
absorption. 
Given that we now have an accurate redshift measurement, we include $\rm SDSS J1335+3533$ into our analysis, which causes the slope of the redshift evolution of the proximity zones to become shallower due to its small zone. We will further discuss the implications of small proximity zones in \S~\ref{sec:small_zones}.

Further differences between the different analyses can be attributed to the different continuum fitting methods. Previous analyses applied a power-law fit to the quasar continuum redwards of the \lya emission line and fitted the \lya and \nv lines with Gaussian curves, whereas we chose to model the continuum of each quasar with two different sets of PCA (i.e. \citet{Paris2011} or \citet{Suzuki2006}). Although differences in the continuum estimation can change the size of individual proximity zones, we do not expect them to alter the distribution of proximity zone sizes, since the scatter on $R_p$ resulting from continuum uncertainties is much smaller than the intrinsic scatter due to density fluctuations \citep{KramerHaiman2009}.

The continuum uncertainties in our analysis arising due to differences between the two sets of PCA are not very significant. The median difference between the proximity zone sizes when estimating the quasar continua with different PCA models is $\langle\Delta R_p\rangle = \langle 
|R_{p, \rm Paris}-R_{p, \rm Suzuki}|\rangle\approx 0.09$~pMpc. The scatter in the distribution of $\Delta R_p$ determined from the $16$th and $84$th percentile is $\sigma_{\Delta R_p}\approx 0.73$~pMpc.

To conclude, it is not completely clear to us, how other authors obtained some very large proximity zones, particularly for a few objects at $z<6$ that are the main drivers for the steep redshift evolution. 
Some of the differences between ours and previous analyses and the resulting shallower redshift evolution of the proximity zone sizes can be attributed to a variety of reasons: first, we stick to a rigorous definition of the proximity zone sizes and continue to take the first drop of the smoothed flux below the $10\%$ level, since we adopt the same rigorous treatment for the mock spectra from the radiative transfer simulations. 
Second, we analyze higher quality data, which can help to determine the drop in flux below the $10\%$ level easier.  
Third, we include all quasars of the ensemble in our analysis and do not exclude the weak emission line quasar in our sample. Finally, we analyze a larger sample of quasar spectra than previous analyses. 
Differences in the continuum estimation, updated redshift and magnitude measurements, and the different correction for the quasar luminosity do not have a significant influence on the discrepancy.

\section{Exceptionally Small Quasar Proximity Zones}\label{sec:small_zones}

Several of the quasars we studied have particularly small proximity zones, as can be seen from Fig.~\ref{fig:Rpvsz}. The proximity zones of two objects, i.e. $\rm CFHQS J2229+1457$ and $\rm SDSS J1335+3533$, are $R_{p} < 1$~pMpc and $R_{p, \rm corr} \lesssim 1$~pMpc. Additionally, given its extreme brightness, the proximity zone of $\rm SDSS J0100+2802$, $R_p=7.12\pm0.13$~pMpc and $R_{p, \rm corr}=3.09\pm0.06$~pMpc, is also exceptionally small. In this section, we discuss the properties of these objects and possible explanations for their small proximity zones. 

\begin{figure*}[t]
\centering
\includegraphics[width=\textwidth]{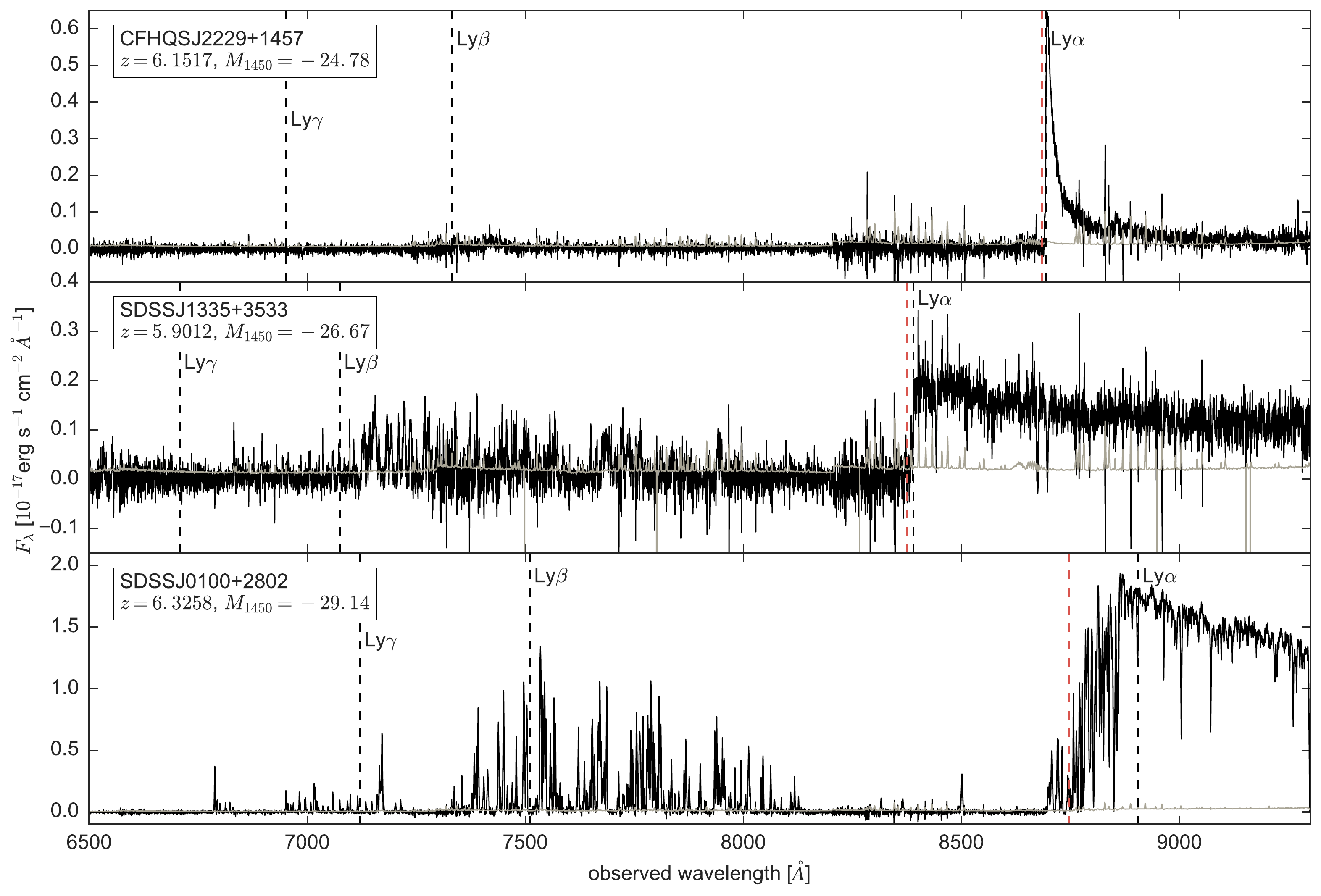}
\caption{Spectra of the three quasars $\rm CFHQS J2229+1457$ (upper panel), $\rm SDSS J1335+3533$ (middle panel) and $\rm SDSS J0100+2802$ (lower panel), exhibiting exceptionally small proximity zones. Ly$\alpha$, \lyb and \lyc lines are indicated by the black dashed lines, whereas the red dashed lines show the extent of the measured proximity zones. \label{fig:spectra_small}} 
\end{figure*}

\subsection{Individual Objects with Small Proximity Zones}

\subsubsection*{CFHQS J2229+1457} 

This relatively faint quasar, $M_{\rm 1450}=-24.78$ \citep{Banados2016}, was discovered in the Canada-France High-z Quasar Survey (CFHQS) and published by \citet{Willott2009}. 
The measurements of the proximity zone are $R_p=0.45\pm0.14$~pMpc and, when normalized to a magnitude of $M_{\rm 1450}=-27$, $R_{p, \rm corr} = 1.07\pm0.33$~pMpc. 
The quasar has a precise redshift measurement of $z=6.1517$ from \cii emission 
from the host galaxy \citep{Willott2015} and hence a very small redshift uncertainty $\Delta v=100$~km/s resulting in an uncertainty of the proximity measurement of $\Delta R_p\approx 0.14$~pMpc. The measurements of the proximity zone are independent of the choice of PCS used to model the quasar continuum, i.e. continuum uncertainties do not influence our measurements.The top 
panel of Fig.~\ref{fig:9spectra_faint}
shows the continuum normalized spectrum and the proximity zone of this
object.
In the upper panel of Fig.~\ref{fig:spectra_small} we show the whole spectrum, revealing no BAL features.

\subsubsection*{SDSS J1335+3533} 

This quasar, that has been discovered in the Sloan Digital Sky Survey (SDSS) by \citet{Fan2004}, has a precise redshift measurement of $z=5.9012$ from CO
$(6-5)$ emission from the host galaxy \citep{Wang2010}, and is
fairly bright with a magnitude of $M_{\rm 1450}=-26.67$
\citep{Banados2016}. The measured proximity zone size for this object
is $R_p=0.78\pm0.15$~pMpc and the luminosity corrected size of the proximity zone is $R_{p, \rm
  corr}=0.89\pm0.17$~pMpc. The redshift uncertainty of $\Delta v=100$~km/s results in an uncertainty of the proximity zone measurement of $\Delta R_p\approx 0.15$~pMpc. These measurements of the proximity zone are again independent on continuum modeling uncertainties. The top 
panel of Fig.~\ref{fig:9spectra}
shows the continuum normalized spectrum and the proximity zone of this
object. 
The whole spectrum of the quasar shown in the middle panel of 
Fig.~\ref{fig:spectra_small} is completely devoid of broad emission
lines, as previously noticed from its discovery spectrum by
\citet{Fan2004}. 
The spectrum does not show any BAL features.

\subsubsection*{SDSS J0100+2802}

This object is the brightest high redshift quasar known so far \citep{Wu2015} with an absolute magnitude of $M_{\rm 1450}=-29.14$ \citep{Banados2016}. Its redshift, $z=6.3258$, has been measured precisely by the detection of the \cii line \citep{Wang2016}. Given its extreme brightness the quasar exhibits a very small proximity zone of $R_p=7.12\pm0.13$~pMpc ($R_{p, \rm corr}=3.09\pm0.06$~pMpc, when normalized to a luminosity of $M_{\rm 1450}=-27$) compared to a proximity zone size of $R_p= 12.0\pm2.0$~pMpc, that one would expect to see for an object this bright. The redshift uncertainty $\Delta v=100$~km/s results in an uncertainty of the proximity zone measurement of $\Delta R_p\approx 0.13$~pMpc. Continuum uncertainties do not play any significant role for these measurements. The lower panel of Fig.~\ref{fig:spectra_small} shows the spectrum of this object, which does not show any BAL features. The continuum normalized spectrum and its proximity zone is shown in the second to last panel of Fig.~\ref{fig:all_spectra} in Appendix~\ref{sec:remaining_zones}. 

\subsection{How Common are Small Proximity Zones?}

\begin{figure*}[t]
\centering
\includegraphics[width=\textwidth]{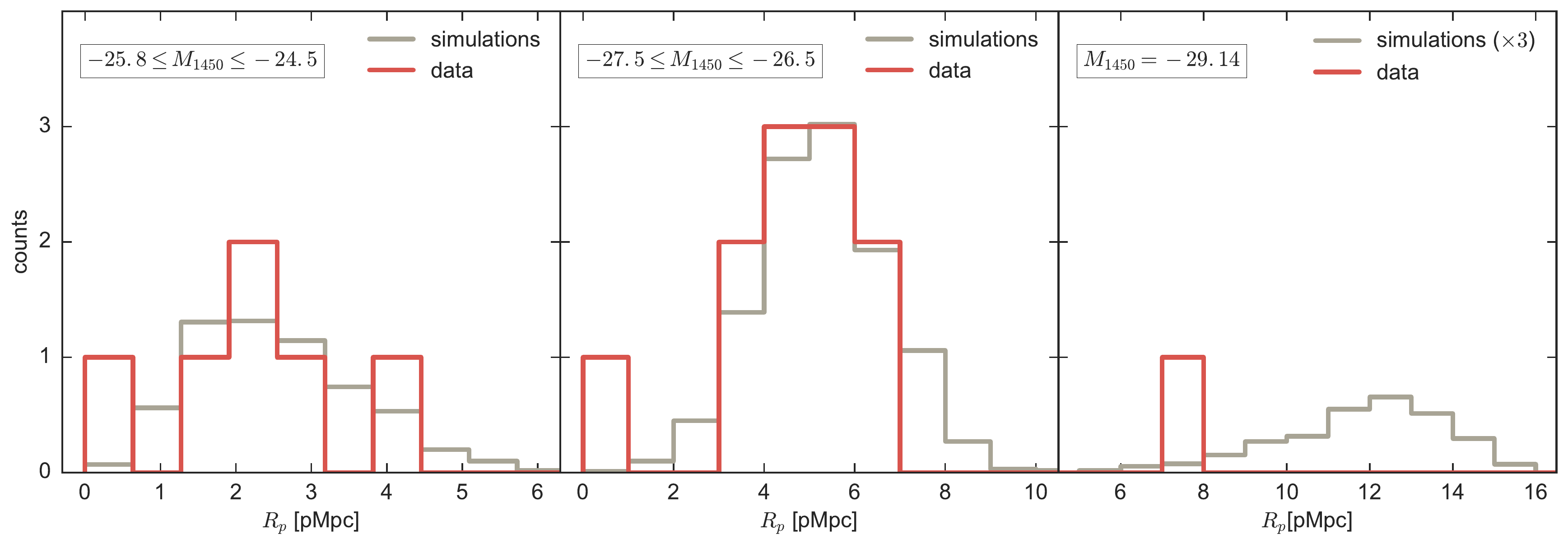}
\caption{Distribution of sizes of quasar proximity zones for faint quasars with luminosities between $-25.8\leq M_{\rm 1450}\leq -24.5$ (left panel), brighter quasars with luminosities between $-27.5\leq M_{\rm 1450}\leq -26.5$ (middle panel) and for very bright quasars with luminosity $ M_{\rm 1450}\approx -29.14$ (right panel). The red histogram shows the distribution of the measured proximity zone sizes of quasars in our data sample within the given luminosity range that have redshift measurements from \mgii, \cii or CO lines. The gray histogram shows the expected distribution for quasars with the same redshifts and luminosity properties as in the data sample from the radiative transfer simulations. In the right panel the simulated distribution is multiplied by a factor of three for better visibility. Our data sample includes six fainter quasars with these requirements, eleven brighter ones and one very bright one. \label{fig:comp_sim}} 
\end{figure*}

In order to quantify the probability of finding objects with such small proximity zones, we compare in Fig.~\ref{fig:comp_sim} our measured
proximity zones $R_p$ from a subset of quasar spectra from our data sample to  proximity zones from our radiative transfer simulations. 
We show the distribution of quasar
proximity zone sizes for three different luminosity ranges, quasars with
magnitudes $-25.8\leq M_{\rm 1450}\leq
-24.5$ are shown in the left
panel, brighter objects with magnitudes $-27.5\leq M_{\rm 1450}\leq -26.5$ are shown on the middle panel, and the right panel shows the distribution of proximity zones for very bright quasars with $M_{\rm 1450}=-29.14$. 
The magnitude intervals were chosen, such that each bin has a width of $\sim 1$~dex and includes one of the small proximity zone objects.
This results in six quasars in the faintest magnitude range, eleven quasars in the brighter range of magnitudes and only one object at $M_{\rm 1450}\approx-29$ \citep{Wu2015}. 
Because of this split into magnitude intervals, we are comparing $R_p$ instead of $R_{p,\rm corr}$, and model the ranges of quasar luminosities in the simulations. 
Note that the redshift range covered by the quasars in each bin is very broad, i.e. $5.844\leq z\leq 6.4323$ for the faintest magnitude bin, $5.7722\leq z\leq 6.5412$ for the brighter magnitude range, and the only one bright quasar has a redshift $z\approx 6.3258$. We forward model the redshift evolution of the quasars in the radiative transfer simulations, such that the broad range of luminosities in each bin does not influence our results.

The red histograms show the distribution of measured proximity zones from our data set. In this figure we only consider quasars that have their redshifts determined by the detection of a \mgii, \cii or CO line, i.e. with small redshift uncertainties. These histograms 
have been normalized to the number of objects available in our data set within the given magnitude interval. 

The gray histograms show the distribution of proximity zone sizes
expected from our radiative transfer runs. We simulate multiple
realizations of each quasar in our sample matching the redshift,
luminosity, and redshift uncertainties (which we add as a Gaussian
uncertainty to the quasar position in our mock spectra; see
Fig.~\ref{fig:mock_spectra}). For the two leftmost panels of Fig.~\ref{fig:comp_sim}
we simulate $100$ skewers for each quasar in our data sample.
Thus the gray histograms show
the proximity zones of $600$ skewers in the left panel and $1100$
skewers in the middle panel. For the single very bright quasar in the right panel of
Fig.~\ref{fig:comp_sim} we simulate
$900$ realizations to improve the statistics.
We do not model continuum uncertainties in the simulations,
which we expect to have negligible influence on the distribution of
proximity zone sizes 
\citep[see discussion in \S~\ref{sec:comparison}, ][]{KramerHaiman2009}. For the radiative transfer
runs we assume a highly ionized IGM, which is consistent with empirical constraints on the neutral gas fraction at $z\sim 6$  \citep{Calverley2011, WyitheBolton2011, Becker2015, McGreer2015}. Following the arguments of
\S~\ref{sec:shallow_evolution} we do not expect any qualitative
changes when assuming a mostly neutral IGM. 

In both the faint and the brighter case (left and middle panel) the
bulk of the distributions of measured and simulated proximity zones
agrees very well. Note that our simulations do not reproduce the
very large proximity zones of $R_p\sim 10-13$~pMpc that
\citet{Carilli2010} measured previously.

However, the simulations also do a poor job of reproducing the frequency of
very small proximity zones that we find. 
For the fainter quasars (left panel) our simulations indicate that the probability of finding a quasar with $R_p\approx 0.45$~pMpc, which is our measurement for $\rm CFHQSJ 2229+1457$, in a sample of six quasar spectra is $\approx 3\%$. 
However, the significance of this small proximity zone is limited by the definition of the proximity zone. The smoothed flux in the spectrum of $\rm CFHQSJ 2229+1457$ remains below the $<10\%$ level within a distance of $10$~pMpc to the quasar, whereas a by eye inspection of the simulated small proximity zones reveals that in most spectra the flux increases above $10\%$ again just outside of their proximity zone. If we would adopt a different definition of the proximity zone that would also be sensitive to the length of the GP trough outside of the proximity zone, this object would be an even greater outlier.

In the middle panel showing quasars with magnitudes $-27.5\leq M_{\rm 1450}\leq -26.5$ we have one object, $\rm SDSS J1335+3533$, with a measured proximity zone $R_p \approx 0.78$~pMpc. In the sample of $1100$ simulated quasars with similar redshifts and magnitudes only one of them has a proximity zone size that small. Thus the probability of finding such an object in a sample of eleven quasars is $\approx 1\%$.

In the right panel we have only one object in our data set, $\rm SDSS J0100+2802$. Its proximity zone is relatively small given its extreme brightness. We estimate the probability of finding a quasar with a proximity zone of $R_p\lesssim 7.12$~pMpc to be $\approx 3\%$. 

In summary, in all three respective magnitude ranges, the occurrence of small proximity zones is much higher in our data sample than the simulations predict. We will now investigate several possible scenarios that could explain the exceptionally small proximity zones of the three aforementioned objects.

\subsection{Possible Explanations for Small Proximity Zones}

We address three scenarios that could possibly explain the exceptionally small proximity zones of the three quasars mentioned in the previous subsection: The proximity zones could be prematurely truncated due to associated dense absorbers, such as damped \lya systems (DLA) or Lyman limit systems (LLS), patches of remaining neutral hydrogen within the IGM could truncate the proximity zones, or the quasars could be very young.

\subsubsection{Truncation Due to Damped \lya Systems or Lyman Limit Systems}

One possible explanation for the exceptionally small proximity zone
sizes 
could be the truncation of the zones due to strong absorbers, such as DLAs
or LLSs, which could either be intervening or
physically associated with the quasar environment (Ba{\~n}ados et al. in prep.). 
The presence of
such self-shielding absorbers could prematurely truncate the proximity
zone at small radii by blocking the quasar's ionizing flux. 
These optically thick absorption line systems are not correctly modeled by our radiative transfer simulations (see \S~\ref{sec:sims}).
We model only the gas densities in the IGM, but do not include such dense gas patches that could be coming from galaxy like overdensities resulting in these absorbers. Thus, proximity zones truncated prematurely by dense absorption systems are not reproduced in our simulations and need to be identified and eliminated from any comparison between the data and the simulations.

To this
end, we search for signatures of strong absorption line systems in the quasar
spectra near the end of their proximity zones by visually inspecting the spectra, and searching for
evidence of damping wings which would indicate the presence of a strong absorber, as well as
associated ionic metal-line transitions.

In the spectrum of $\rm SDSS J1335+3533$, the only strong metal line
absorber we could find was a \mgii absorber at $z\approx2.10$.
However, we have to note that the signal-to-noise ratio of the
spectrum is insufficient to identify very weak metal absorption features.

\begin{figure}[t]
\centering
\includegraphics[width=.5\textwidth]{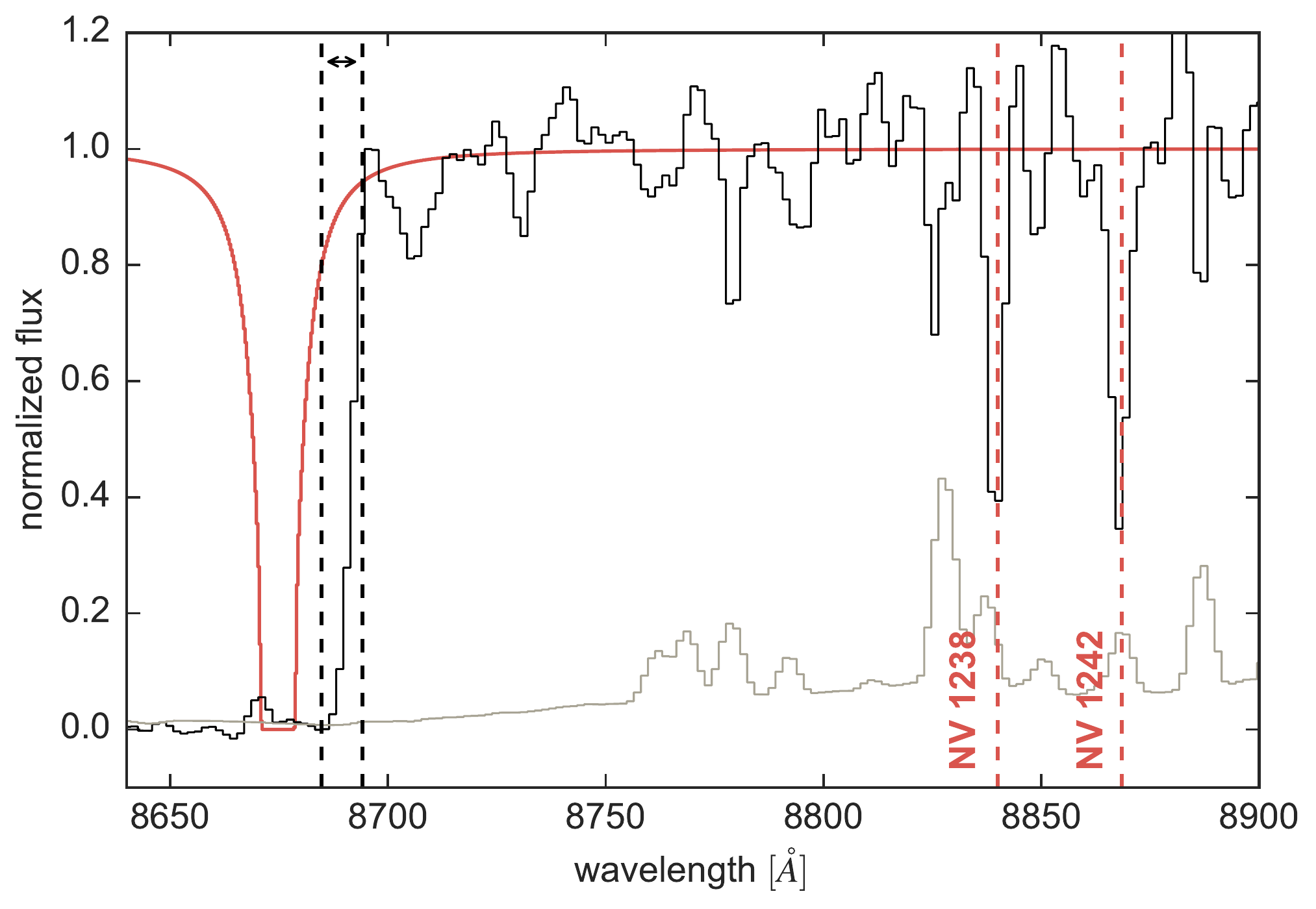}
\caption{Excerpt of the spectrum of $\rm CFHQS J2229+1457$ taken with LRIS. We find an \nv absorption line system (red dashed lines) associated with an absorber at $z\approx 6.136$ (red curve) with a column denisty $N_{\rm HI}=10^{19}\rm cm^{-2}$ and Doppler parameter $b=40$~km/s. A boxcar smoothing of two pixels has been applied to both the spectrum (black curve) and its noise vector (gray curve). The vertical black dashed lines indicate the extent of the proximity zone. \label{fig:abs2229}} 
\end{figure}

Searching the ESI spectrum of $\rm CFHQS J2229+1457$ for
metal absorption lines associated with nearby dense absorption
systems did not reveal any absorbers due to the very low
signal-to-noise data ($\rm S/N\approx2$ per pixel). 
To facilitate the search for
metal lines, we obtained a higher quality spectrum with the Low
Resolution Imaging Spectrometer (LRIS) at the Keck I telescope, which
has a lower resolution ($R\sim 1800$) than the ESI spectrum ($R\sim 4000$) but a 
higher signal-to-noise ratio ($\rm S/N\approx7$ per pixel). 
An excerpt of this spectrum is shown in Fig.~\ref{fig:abs2229} and reveals an \nv doublet associated with an
absorber at $z\approx 6.136$, whose \ion{H}{1} absorption Voigt profile is shown as the red curve when assuming a column density of $N_{\rm HI}=10^{19}$~cm$^{-2}$ and a Doppler parameter of $b=40$~km/s. The absorber is located at a distance of $0.91$~pMpc from the quasar and thus it is highly unlikely that this absorber influences the size of the proximity zone significantly, because the
proximity zone ends at a higher redshift of $z\approx6.144$ at a distance $R_p\approx 0.45$~pMpc. 
Additionally, at the location of the absorber in the spectrum, there is a transmitted flux spike in the \lya forest partly visible (at $\lambda\approx 8670$~{\AA}), which indicates that the absorber cannot be saturated, implying $N_{\rm HI}\lesssim10^{14}$~cm$^{-2}$, 
i.e. its column density would need to be much less than the here assumed $N_{\rm HI}=10^{19}$~cm$^{-2}$ in order not to violate the spectrum, and is thus unlikely to truncate the proximity zone. We also do not find evidence for any low-ion absorption lines, such as \siii at $\lambda_{\rm rest}=1260.42$~{\AA}, that should be present in DLAs.  

\begin{figure*}[t]
\centering
\includegraphics[width=\textwidth]{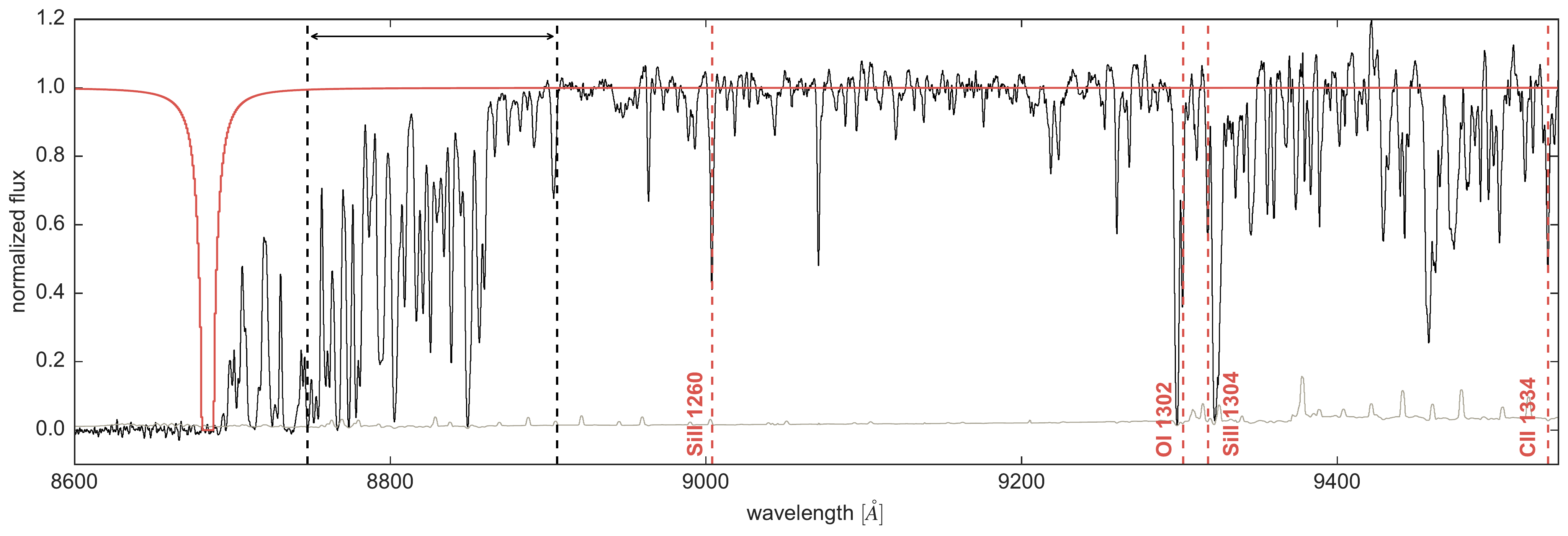}
\caption{Excerpt of the spectrum $\rm SDSS J0100+2802$. We find a close low ionization absorption system at $z\approx 6.144$ (red curve) with column density $N_{\rm HI}=10^{19}\rm cm^{-2}$ and Doppler parameter $b=40$~km/s and associated absorption lines (red dashed lines). The continuum normalized\footnote{Note that in this figure we show the quasar spectrum normalized with a hand-fitted continuum.} spectrum (black curve) and its noise vector (gray curve) are smoothed with a three pixel boxcar function. The vertical black dashed lines indicate the extent of the proximity zone. \label{fig:abs0100}} 
\end{figure*}

In the spectrum of $\rm SDSS J0100+2802$ we find a low ionization absorption system close to the object at $z\approx 6.144$ (see Fig.~\ref{fig:abs0100}), which has previously been identified by \citet{Wu2015}. However, this absorber is at such large
line-of-sight distance $R\approx9.94$~pMpc from the quasar that it cannot truncate its proximity zone, that
already ends at a distance of $R_p=7.12\pm0.13$~pMpc.
We show the \ion{H}{1} Voigt absorption profile (red curve) assuming an absorber with $N_{\rm HI}=10^{19}$~cm$^{-2}$ and $b=40$~km/s, and its associated lines (\siii at $\lambda_{\rm rest}=1260.42$~{\AA} and $\lambda_{\rm rest}=1304.37$~{\AA}, \oi at $\lambda_{\rm rest}=1302.16$~{\AA}, \cii at $\lambda_{\rm rest}=1334.53$~{\AA}).
Note that an absorption system with a higher column density of $N_{\rm HI}\sim10^{20.3}$~cm$^{-2}$ characteristic
of DLAs, would be too broad and would be inconsistent with the presence of flux transmission 
that are observed
in the \lyb forest at this redshift (see \S~\ref{sec:neutral_islands}).

In light of the possibility that proximate absorbers could truncate the proximity zones, it is interesting 
 to estimate the number of those proximate absorbers that one would expect to find around quasars. 
Therefore, we start with an estimate of intervening absorbers along the line-of-sight to quasars by \citet{SongailaCowie2010}, who have estimated the number
density $\diff N/\diff z$ of intervening LLSs in Fig.~$5$ and Table~$3$ of their paper. 
They determine the best power-law fit to their measurements with a maximum likelihood analysis to be
\begin{align}
\frac{\diff N}{\diff z}\approx 2.84\left(\frac{1+z}{4.5}\right)^{2.04}, 
\end{align}
which results in an estimate of $\diff N/\diff z\approx 6.99$ at $z=6$.

On the one hand the abundance of LLSs could be enhanced in the proximity of quasars 
because of the overdense quasar environment \citep{Hennawi2006, Prochaska2008, QPQ6_2013}. However, the intense radiation from the quasar will also ionize the
dense gas in its surroundings making it less likely to self-shield,
possibly lowering the abundance of so-called proximate LLSs \citep{HennawiProchaska2007, Prochaska2008}. \citet{Prochaska2010} found empirically that at $z\sim 4$ the number density of proximate LLSs (occurrence within $\Delta v \leq 3000$~km s$^{-1}$ of the quasar emission redshift) roughly equals
the number density of intervening LLSs \citep[see Fig.~$15$ of][]{Prochaska2010}.
Assuming the same approximate equality holds at $z\sim 6$,
we estimate the probability $p$ of
finding a proximate LLS within $1$~pMpc to a quasar at $z=6$ to be
$p\approx n(z)\Delta z\approx0.11$. 
Given the size of our data set, which consists out of $30$ quasars, 
we would expect to find a proximate LLS
in $\sim 3$ spectra. 

Where are the $\sim 3$ proximate LLSs that we expect to find in our data sample? We found only one in the spectrum of the quasar $\rm SDSS J0840+5624$, which we excluded from our analysis for this reason (see Appendix~\ref{sec:J0840}). The absence of other proximate LLSs detected in our data sample could be attributed to a few different reasons: the estimate of proximate LLSs by \citet{Prochaska2010} assumes a $\Delta v \leq 3000$~km s$^{-1}$ window, which corresponds to a distance of $\approx 4.3$~pMpc to a quasar at $z=6$. We are just interested in the innermost $1$~pMpc to the quasar, where the quasar's radiation is stronger and could thus have photoionized all possible LLSs \citep{HennawiProchaska2007}. A second possible reason for the lack of detections of more proximate LLSs in our sample, could be due to the fact that our data sample at $z\sim 6$ is brighter than the one analyzed by \citet{Prochaska2010} at $z\sim 3.5$ 
and thus the higher radiation could have photoionized more LLSs. Hence, it might be not surprising, that we are only detecting one proximate LLS in our data sample instead of the expected $\sim 3$.

Although we cannot rule out the presence of absorbers with weak metal lines or rare metal-free absorbers \citep{Fumagalli2011, Simcoe2012, Cooper2015, Crighton2016, Cooke2017} in all objects, which would require higher quality data, we do not see evidence for proximate LLSs in the three spectra showing small proximity zones, particularly in the spectrum of the very bright quasar $\rm SDSS J0100+2802$, where we would most likely be able to detect it. 
Follow up higher $\rm S/N$ and resolution observations are required to confirm the lack of associated dense absorbers in these spectra.

\begin{figure*}[t]
\centering
\includegraphics[width=\textwidth]{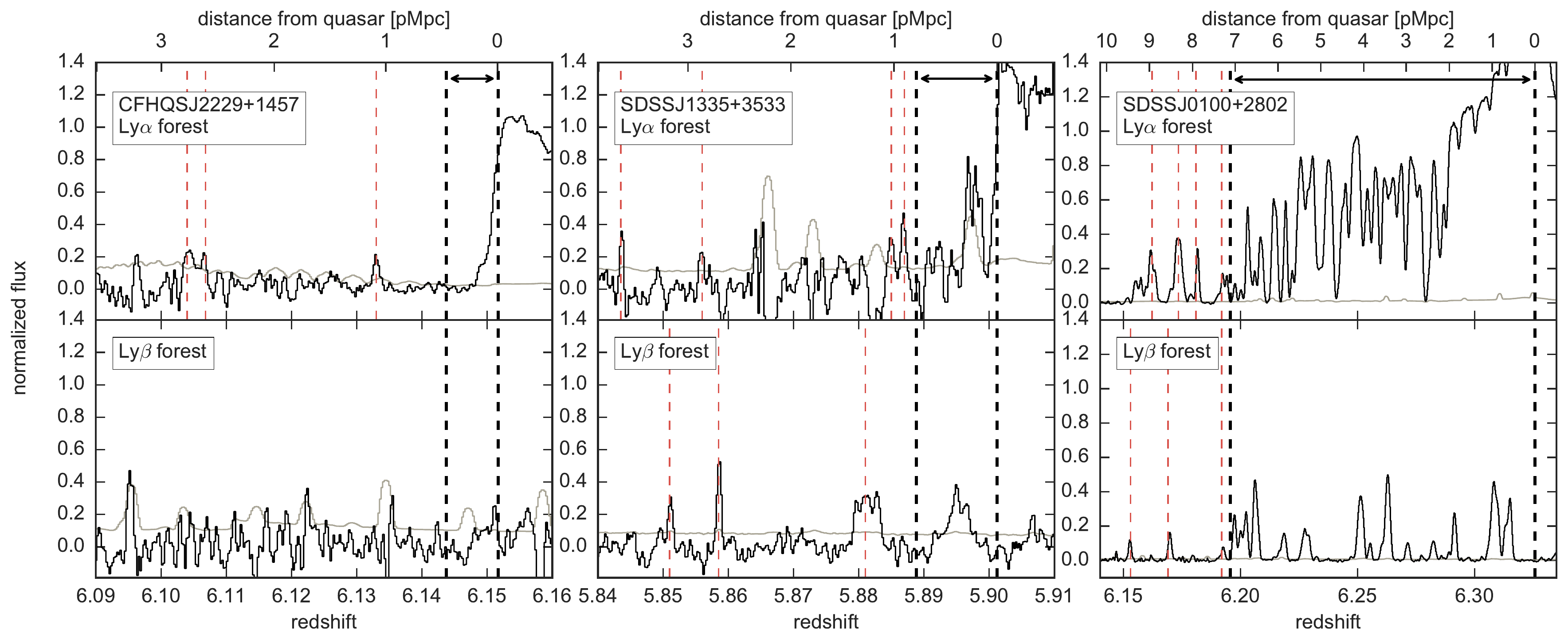}
\caption{Excerpts of the \lya and the \lyb forests of the three spectra of $\rm CFHQSJ 2229+1457$ (left panels), $\rm SDSSJ 1335+3533$ (middle panels) and $\rm SDSS J0100+2802$ (right panels). The panels show the \lya and the \lyb forest (upper and lower panels, respectively) of the spectra at the same redshift and distance close to the quasars. A boxcar smoothing of three pixels has been applied to both the spectra (black curves) and noise vectors (gray curves). The extents of the proximity zones are indicated by the vertical black dashed lines. Red dashed lines indicate transmission spikes outside of the respective proximity zones.  \label{fig:lyab}} 
\end{figure*} 

\begin{figure*}[t]
\centering
\includegraphics[width=\textwidth]{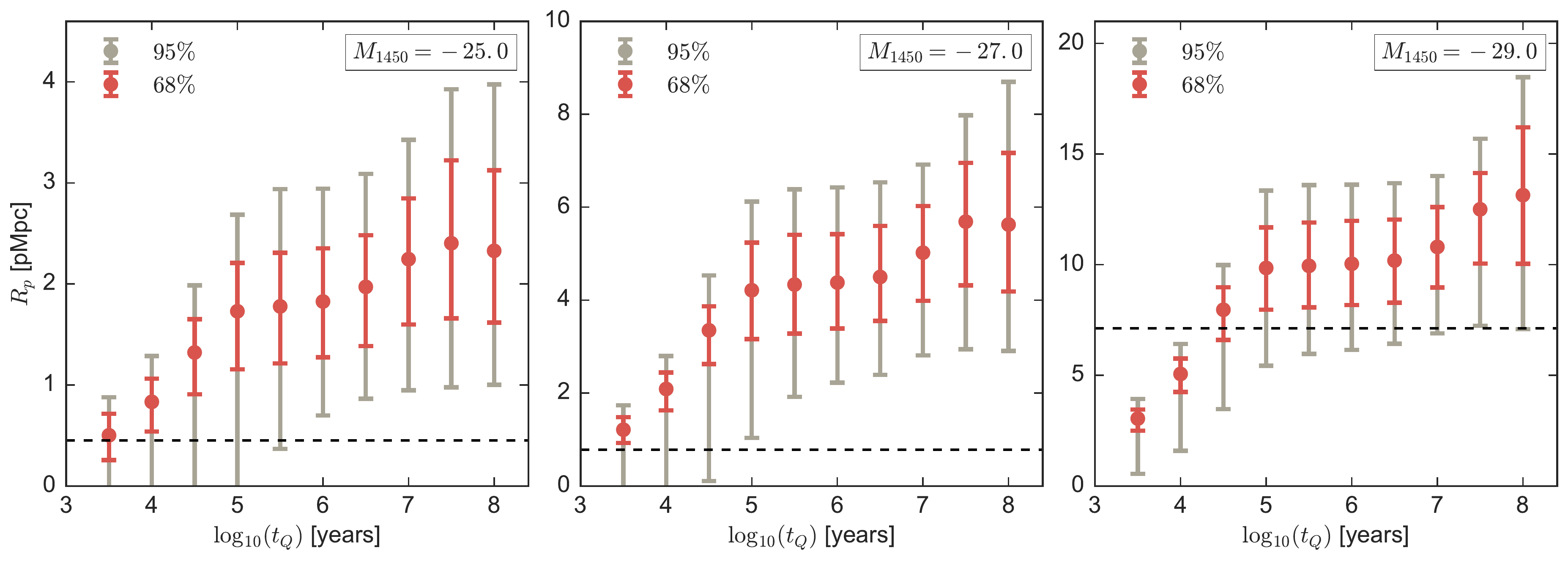}
\caption{Distributions of proximity zone sizes shown as a function of the average quasar age $t_{\rm Q}$ for quasars with magnitudes of $M_{\rm 1450}=-25.0$ (left panel), magnitudes of $M_{\rm 1450}=-27.0$ (middle panel) and magnitudes of $M_{\rm 1450}=-29.0$ (right panel). The proximity zone sizes of $\rm CFHQS J2229+1457$, $\rm SDSS J1335+3533$ and $\rm SDSS J0100+2802$ are shown as the dashed black line in the left, middle and right panel, respectively. \label{fig:lifetime}} 
\end{figure*} 

\subsubsection{Neutral Islands of Gas in a Patchily Reionized IGM}\label{sec:neutral_islands}

Another possible explanation for the existence of very small proximity zones could be the truncation of the proximity
zones due to islands of very neutral gas in an otherwise highly ionized IGM, which can occur in 
patchy reionization models \citep[e.g.][]{Lidz2007, Mesinger2010}. 
This scenario implies that reionization is not yet complete at $z\sim6$ and these quasars lie in (presumably rare) neutral environments that could absorb all incoming flux and thus truncate the proximity zones. Opposite to the previously described scenario where LLSs, i.e. dense gas on galactic scales, truncate the proximity zones, we assume here neutral patches of $\sim 10$~pMpc scales that are associated with reionization and do not have any low redshift counter parts. 
This explanation requires
the presence of large patches of highly neutral gas just outside the proximity zone,
which would also absorb flux in the Lyman-series (\lya, \lyb, etc.) forests. In other words, in a
patchy reionization scenarios one expects to see long GP troughs that are line black in both \lya
and \lyb, beginning right at the edge of the proximity zone.

In Fig.~\ref{fig:lyab} we show continuum normalized spectra of the
\lya and the \lyb forest (top and bottom panels, respectively) near the three quasars in question. The black dashed lines in all panels indicate the extent of the quasar proximity zone. 
The left panels show the \lya and the \lyb
forest close to the quasar $\rm CFHQS J2229+1457$. While we do not
detect any transmitted
flux in the \lyb forest, we do see a few transmission spikes in the
\lya forest at $R\approx 1.1$~pMpc and $R\approx 2.8$~pMpc, indicated by the red dashed lines. The
transmission in the \lyb forest at the same distances must have been
absorbed 
due to foreground \lya absorption at $z\approx 5.02$ and
$z\approx 4.99$. 

The middle panels show the \lya and \lyb forest of $\rm SDSS J1335+3533$, where we see
transmitted flux in the \lya forest just outside of the proximity zone, i.e. at $R\approx 0.9$~pMpc and
$R\approx 1.0$~pMpc, and a bit further away at $R\approx 2.9$~pMpc and
$R\approx 3.6$~pMpc.
In the \lyb forest we can see clear transmission
spikes just outside of the proximity zone at $R\approx 1.4$~pMpc, $R\approx 2.8$~pMpc and $R\approx 3.3$~pMpc.

The right panels show excerpts of the \lya and \lyb forest for $\rm SDSS J0100+2802$. One can clearly see several transmitted flux spikes in the \lya forest right outside 
the proximity zone. Also the \lyb forest shows a few flux spikes close to the proximity zone, although less prominent ones.

These detections of transmitted flux just outside the proximity zones in the \lya or \lyb forest suggest that if there are patches of neutral gas present close to the quasar, they need to be $\lesssim1$~pMpc in size, corresponding to the size of the regions that are both line black in the \lya and \lyb forest.
Although we cannot completely rule out a patchy reionization scenario with such small islands of neutral gas, our constraints on their sizes
already suggest that they would be difficult to distinguish from metal poor LLSs.  Deeper spectra (at least for the quasars $\rm CFHQS J2229+1457$ and $\rm SDSS J1335+3533$) covering their proximity zones further down the Lyman series could reveal additional \lyb or \lyc transmission spikes that would further
constrain this scenario, as well as enabling more sensitive searches for metal lines to rule out an LLS scenario.

\subsubsection{Young Quasar Age}\label{sec:lifetime}

Finally, another possibility to explain small proximity zones is that the quasars may have turned on only very recently. 
When a quasar turns on the IGM requires a finite amount of time to
adjust to the new higher photoionization rate.  We will focus the
discussion in this section on a highly ionized surrounding IGM,
consistent with optical depth measurements at $z\sim 6$
\citep[e.g.][]{WyitheBolton2011, Calverley2011, Becker2015,
  McGreer2015}, although we note similar arguments apply in the
neutral case \citep[see][for further details]{Khrykin2016}. Following \citep{Khrykin2016}\footnote{\citet{Khrykin2016} wrote down the analogous equation for the \ion{He}{2} singly ionized fraction in the context of \ion{He}{2} Ly$\alpha$ proximity zones at $z\sim 3$, the exact same set of arguments applies here for neutral hydrogen around $z\sim 6$ quasars.} 
the time evolution of the neutral fraction in the proximity
zone is well described by the equation 
\begin{align}
x_{\rm HI}(t_{\rm Q}) = x_{\rm HI,0} + (x_{\rm HI,0}-x_{\rm HI,eq})\exp^{-t_{\rm Q}\slash t_{\rm eq}} \label{eq:HI}, 
\end{align}
where $t_{\rm Q}$ is the quasar age, $x_{\rm HI,0} \approx n_{\rm e}\alpha_{\rm HII}\slash \Gamma_{\rm UVB}$ is the neutral
fraction of the IGM before the quasar turned on, and $x_{\rm HI,eq} \approx n_{\rm e}\alpha_{\rm HII}\slash (\Gamma_{\rm UVB} + \Gamma_{\rm QSO})$
is the new lower neutral fraction that the IGM will reach once ionization equilibrium is attained.  Here $n_{\rm e}$ and $\alpha_{\rm HII}$ are the IGM electron
density and recombination coefficient, respectively. The characteristic timescale to reach ionization equilibrium, the equilibration timescale $t_{\rm eq}$, 
is
\begin{equation}
t_{\rm eq}\approx \frac{1}{\Gamma_{\rm UVB} + \Gamma_{\rm QSO}}. 
\end{equation}
Eqn.~(\ref{eq:HI}) thus implies that a quasar has to have been shining for at least $t_{\rm Q}\gtrsim t_{\rm eq}$, in order for the surrounding IGM
to have reached ionization equilibrium with the quasar and result in the maximum proximity zone size. Ionizing equilibrium is achieved more rapidly close the quasar due to the stronger radiation field ($\Gamma_{\rm QSO} \propto R^{-2}$).
In order to obtain an optical depth of $\tau_{\rm lim}=2.3$ at $z\sim 6$, which corresponds to a transmitted flux level of $10\%$ in the \lya forest, a value of $\Gamma_{\rm QSO} \sim 10^{-12}$~$\rm s^{-1}$ is implied (see \S~\ref{sec:shallow_evolution}),
resulting in a typical equilibration time scale of $t_{\rm eq}\sim 3 \times 10^4$~yr. Full ionization equilibrium is reached after a few
equilibration times, or for $t_{\rm Q}\sim 10^5$~yr.
The region in which ionization equilibrium is reached, and hence also the size of 
the quasar's proximity zone, will  grow with time until $t_{\rm Q}\approx 3t_{\rm eq}$. Thus, even for a highly ionized IGM,
the sizes of the
proximity zones are dependent on the quasar age 
for $t_{\rm Q} \lesssim 10^5$~yr, whereas for ages $t_{\rm Q} \gtrsim 10^5$~yr the proximity zone sizes are largely independent on the precise quasar age\footnote{A mild increase in proximity zone sizes for long quasar ages, i.e. $t_{\rm Q}\gtrsim 10^7$~yr, can be attributed to heating effects due to \heii reionization \citep{Bolton2010, Bolton2012, Daviesinprep}}.

In Fig.~\ref{fig:lifetime} we show the dependence of proximity zone size on quasar age determined from
radiative transfer simulations
in a highly ionized IGM. Quasars with magnitudes $M_{\rm 1450}=-25.0$ (left) $M_{\rm 1450}=-27.0$ (middle) and $M_{\rm 1450}=-29.0$ (right) are shown,
where we have simulated $900$ proximity zones at each magnitude.
The black horizontal dashed lines show the measured sizes of the proximity zones of $\rm CFHQS J2229+1457$ (left panel), $\rm SDSS J1335+3533$ (middle panel) and $\rm SDSS J0100+2802$ (right panel).

The distribution of proximity zone sizes in the left and middle panel encompasses values of
$R_p\approx 1$~pMpc that we have measured only when assuming a short quasar age of $t_{\rm Q}\lesssim 10^5$~yr.
Hence a plausible explanation for the
exceptionally small sizes of the proximity zones could be young ages
for these quasars. 
The proximity zone of the very bright quasar $\rm SDSS J0100+2802$ (right panel) is not as significant of an outlier 
as the other two objects, but nevertheless the small size of its zone would be more probable for $t_{\rm Q} \sim 10^5$~yr.

At lower redshifts $z\lesssim 4$, studies of quasar clustering imply that the duty cycle of quasar activity is $t_{\rm dc}\sim10^9$~yr \citep{Shen2007, WhiteMartiniCohn2008}\footnote{Whether this also applies at redshifts $z\sim 6$ still remains to be determined. }. However, whereas clustering constraints the
duty cycle, proximity zones actually probe a different timescale, which is the duration of quasar emission episodes which could be considerably shorter. Indeed, we actually obtain a lower limit on the episodic lifetime, because the quasar could continue to emit for many years after we observe it on Earth. 
Nevertheless, for a given episodic lifetime $t_{\rm episodic}$ the probability of measuring an age of $t_{\rm Q}$ is $p=t_{\rm Q}\slash t_{\rm episodic}$,
assuming the simplest ``light-bulb'' lightcurve for the quasars. 
If we assume an average episodic lifetime of $t_{\rm episodic}\sim 10^8$~yr, the probability of observing a quasar that has only been shining for $\sim 10^5$~yr is
$p\sim 0.1\%$, or $p\sim 1\%$ for $t_{\rm episodic}\sim 10^7$~yr. We have discovered three objects suggesting ages of $t_{\rm Q}\sim 10^5\,{\rm yr}$
in a sample of $30$ quasars, i.e. $p\approx 10\%$. To be consistent with finding a few of these small proximity zones, the quasar episodic lifetimes would
need to be $t_{\rm episodic}\sim 10^6$~yr. This would leave the sizes of the proximity zones of the vast majority of quasars unchanged, which we typically observe much later in their evolution, but could explain the very small zones we find.

\section{Summary}\label{sec:summary}

In this paper we analyze a sample of $34$
high redshift quasar spectra taken with the ESI instrument on the Keck II telescope. 
We reduce the spectra in a homogeneous way, and analyze the sizes of their proximity zones for a subset of $30$
quasars which do not exhibit BAL features or have obvious
signatures of nearby associated absorption line systems that could prematurely
truncate their proximity zones. Our analysis uses updated redshift measurements,
fully consistent values for the quasar 
absolute magnitudes $M_{\rm 1450}$, and carefully fits to the quasar continua based on
principal component spectra.

Previous work found a strong evolution of proximity zone sizes with redshift,
and it was argued that this provided evidence for rapid evolution of the IGM neutral fraction during the epoch of reionization. 
We instead find a much shallower redshift evolution, which is however consistent with
the evolution predicted by our radiative transfer simulations, irrespective of assumptions about the ionization state of the IGM. 
The size of the proximity zone ends at a distance corresponding to a flux transmission level of $10\%$ according to the definition of quasar proximity zones \citep{Fan2006} and, in a highly ionized IGM, this distance is reached at a location where the ionization rate of the quasar dominates the ionization rate of the UVB by at least an order of magnitude.  As such, the size of the quasar proximity zone is essentially independent of the UVB and neutral gas fraction. Assuming a neutral ambient IGM the observed proximity zone sizes $R_p$  track the growth of the ionized \hii region $R_{\rm ion}$, but then cease to grow
when the distance with a $10\%$ flux transmission level is reached. For a highly plausible quasar age of $t_{\rm Q}\sim 10^{7.5}$~yr, $R_p$ is thus independent of the ionization state of the IGM for a mostly neutral IGM as well. Thus contrary to previous claims, both the observed shallow redshift evolution and
the results from our simulations imply that the redshift evolution of proximity zone size $R_{p}$ does not significantly constrain the evolution of the neutral gas fraction during the epoch of
reionization.

Our analysis uncovered three quasars with exceptionally small
proximity zones, two of them with $R_p\lesssim 1$~pMpc. The occurrence of zones this small is extremely
rare in our simulations. We estimate the probability of finding these objects  in a sample of $30$ quasar spectra
to be $p\approx 1\%$ for $\rm SDSS J1335+3533$, and $p\approx
3\%$ for $\rm CFHQS J2229+1457$ and $\rm SDSS
J0100+2802$. We search for evidence of proximate absorption systems such as DLAs or LLSs, or islands of remaining neutral hydrogen associated with patchy reionization, both of which could result in truncation of the proximity zones closer to the quasars.  However the absence of
strong metal absorption lines or large GP troughs (in both the \lya and the \lyb forest) appears to disfavor 
these truncation scenarios. Nevertheless we cannot completely rule out the presence of low metallicity or metal-free absorbers due to the low signal-to-noise ratio of the data of $\rm CFHQS J2229+1457$ and $\rm SDSS J1335+3533$.

The most plausible explanation for the small proximity zones that we observe is that these quasars are relatively young.
It takes the IGM roughly $\sim 10^5\,{\rm yr}$ to reach ionization equilibrium with the quasar radiation.
Proximity zones of quasars with ages longer than this timescale are essentially independent of the exact age of the quasar. However, for shorter
quasar ages the surrounding IGM has not yet reached ionization equilibrium resulting in proximity zones comparable to the small sizes we observe.
If the duration of quasar emission episodes is $t_{\rm episodic}\sim 10^8$~yr, the detection of
these small zones would be very unlikely, i.e. $p\sim 0.1\%$. This discrepancy can be resolved if one
assumes a shorter duration of $t_{\rm episodic}\sim 10^6$~yr, resulting in a
probability of $p\sim10\%$ for finding these objects with ages of
$t_{\rm Q}\sim 10^5$~yr. 

However, an average episodic lifetime of $t_{\rm episodic}\sim 10^6$~yr causes significant tension with the inferred sizes of
SMBHs at these redshifts, since the presence of $\sim 10^{9}M_\odot$ SMBHs at $z\sim6$ requires that quasars accrete
continuously at around the Eddington limit for nearly the entire Hubble time of $\sim 10^9{\rm yr}$ \citep{Volonteri2012}. 
Thus, although the proximity zone measurements constrain the episodic lifetime $t_{\rm episodic}$, whereas the sizes of SMBH depend
on the integral over all emission episodes, i.e. the duty cycle, the presence of large SMBHs at  $z\sim 6$ either requires long episodic lifetimes
comparable to the duty cycle $\sim 10^9\,{\rm yr}$, or that the quiescent time between episodic emission bursts must be very short, implying that SMBHs grow via many short episodic phases \citep{Schawinski2015, MaoKim2016, Daviesinprep}.

It would clearly be interesting to uncover more quasars with small proximity zones, particularly if they are
indeed young quasars that have only been shining for $t_{\rm Q}\sim 10^5\,{\rm yr}$.  In the future, we plan to use large quasar surveys,
such as Pan-STARRS or SDSS, to identify candidate small-zone quasars and measure accurate redshifts from \cii or CO emission lines with the Atacama Large Millimeter Array (ALMA), in order to confirm the potentially small sizes of their proximity zones. Deep optical and near-IR follow-up observations will be helpful to rule out a premature truncation due to associated absorption systems. 
Further theoretical work is also required to investigate the causes and implications of small proximity zones, and their relationship to the
distribution of quasar ages, which is a subject we address in future work \citep{Daviesinprep}.

\section*{Acknowledgment}
The authors would like to thank J. Bolton and M. Haehnelt for their questions about the small proximity zones in our sample at the
``Dark Ages'' meeting in Heidelberg, which provided part of the
motivation for this work.  Additionally, we wish to thank E. Ba{\~n}ados, X. Fan, B. Venemans, and F. Walter for valuable input and discussion. 

The data presented in this paper were obtained at the W.M. Keck
Observatory, which is operated as a scientific partnership among the
California Institute of Technology, the University of California and
the National Aeronautics and Space Administration. The Observatory was
made possible by the generous financial support of the W.M. Keck
Foundation.

This research has made use of the Keck Observatory Archive (KOA), which is operated by the W. M. Keck Observatory and the NASA Exoplanet Science Institute (NExScI), under contract with the National Aeronautics and Space Administration.

The authors wish to recognize and acknowledge the very significant cultural role and reverence that the summit of Mauna Kea has always had within the indigenous Hawaiian community.  We are most fortunate to have the opportunity to conduct observations from this mountain. 

ZL was supported by the Scientific Discovery through Advanced Computing (SciDAC) program funded by U.S.~Department of Energy Office of Advanced Scientific Computing Research (ASCR) and the Office of High Energy Physics. Calculations presented in this paper used resources of the National Energy Research Scientific Computing Center (NERSC), which is supported by the Office of Science of the U.S. Department of Energy under Contract No. DE-AC02-05CH11231.  The authors would like to thank Dmitriy Morozov and Gunther H.~Weber for the halo finder work, which was funded by the ASCR project ``Scalable Analysis Methods and In Situ Infrastructure for Extreme Scale Knowledge Discovery,'' program manager Lucy Nowell.

\bibliography{literatur_hz}

\appendix

\section{Associated Absorbers in SDSSJ0840+5624}\label{sec:J0840}

This object has been excluded from our analysis because we found several absorption systems at high redshift associated with the quasar itself. 
These absorbers cause additional absorption within the proximity zone of the quasar, such that our analysis recovers a spuriously small proximity zone of $R_p\approx 0.88$~pMpc. However, inspecting the spectrum by eye reveals that the actual proximity zone of this object extends much further out to $R_p\gtrsim 5.0$~pMpc. 

\begin{figure}[h]
\centering
\includegraphics[width=.5\textwidth]{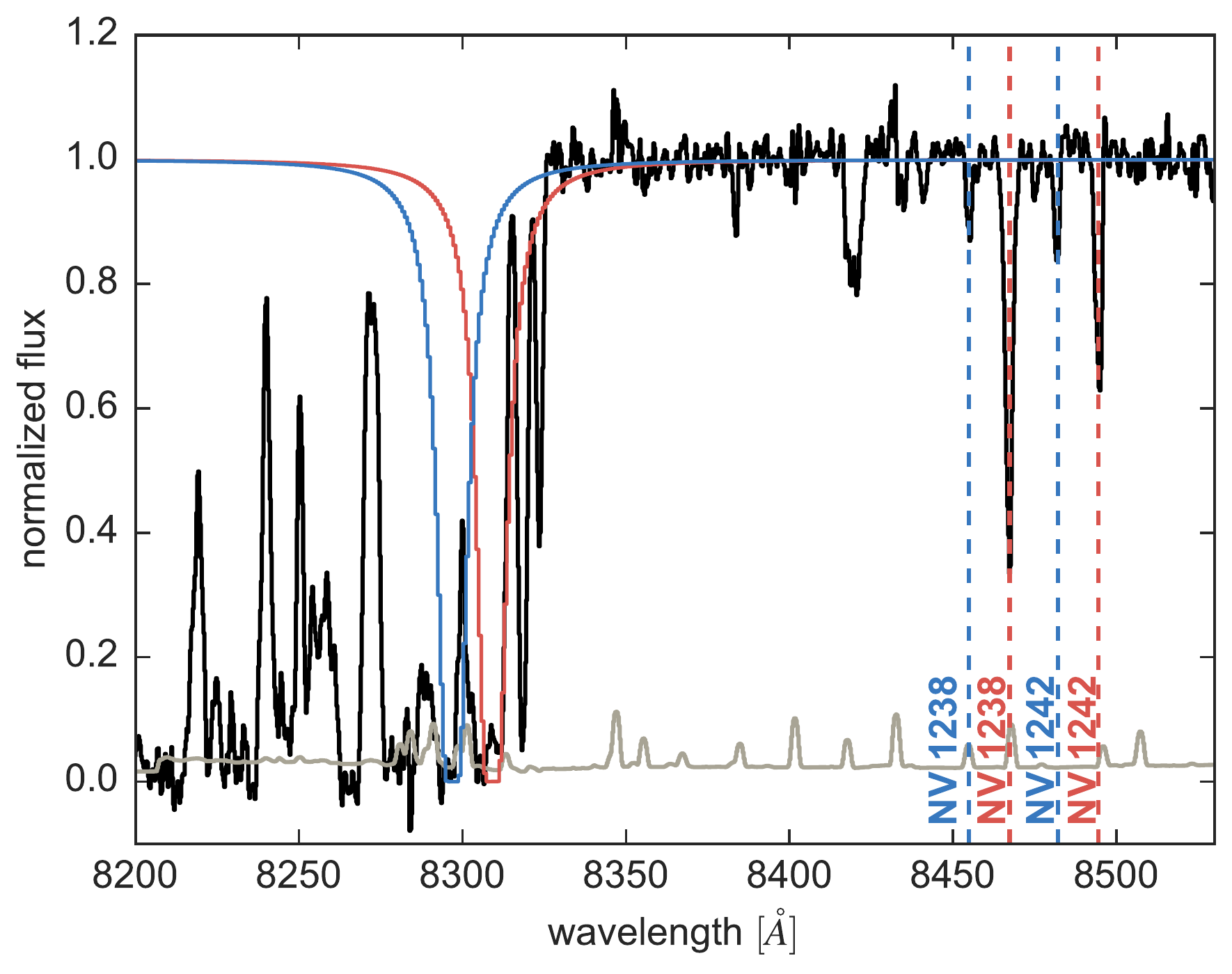}
\caption{Two \nv absorption line systems shown as the blue and red dashed lines asscoiated with close absorbers (blue and red curves) at $z_{\rm blue}\approx 5.825$ and $z_{\rm red}\approx 5.835$ in the spectrum of the quasar $\rm SDSSJ0840+5624$. The continuum normalized\footnote{Note that in this figure we show the quasar spectrum normalized with a hand-fitted continuum.} spectrum (black curve) and its noise vector (gray curve) are smoothed with a three pixel boxcar function. The depicted absorbers have a column density $N_{\rm HI}=10^{19}\rm cm^{-2}$ and a Doppler parameter $b=20$~km/s.  \label{fig:abs0840}} 
\end{figure}

We find two \nv absorption line systems ($\lambda_{\rm rest}=1238.82$~{\AA} and $\lambda_{\rm rest}=1242.80$~{\AA}) associated with close absorbers at $z\approx5.835$ and $z\approx5.825$, which are shown in Fig.~\ref{fig:abs0840}. We also find another absorption system in the spectrum at $z\approx5.594$ with associated \siii absorption ($\lambda_{\rm rest}=1260.42$~{\AA}) that falls within the proximity zone. Thus there is a lot of absorption within the proximity zone that cannot be contributed to residual neutral hydrogen only and an exact determination of the proximity zone for this object without contamination from other absorption systems is impossible. 

\clearpage
\section{Details of the Continuum Modeling of Each Quasar}\label{sec:details_cont}
Table~\ref{tab:continuum} shows the details of modeling the continuum of each quasar spectrum. 

\begin{deluxetable*}{@{\extracolsep{\fill}}lccc@{}}
\tablecaption{Details of the continuum modeling of each quasar. \label{tab:continuum}}
\tablehead{\colhead{object} & \colhead{normalization to unity at rest-frame} & \colhead{set of PCS} &  \colhead{number of PCS}}
\startdata
SDSS J0002+2550 & $1280$~{\AA} & \citet{Suzuki2006} & $5$ \\
SDSS J0005-0006 & $1280$~{\AA} & \citet{Paris2011} & $5$ \\
CFHQS J0050+3445 & $1280$~{\AA} & \citet{Paris2011} & $5$ \\
SDSS J0100+2802 & $1280$~{\AA} & \citet{Paris2011} & $5$ \\
ULAS J0148+0600 & $1280$~{\AA} & \citet{Paris2011} & $5$ \\
CFHQS J0210-0456 & $1245$~{\AA} & \citet{Paris2011} & $5$ \\
PSO J0226+0302 & $1280$~{\AA} & \citet{Paris2011} & $5$ \\
CFHQS J0227-0605 & $1265$~{\AA} & \citet{Paris2011} & $5$ \\
SDSS J0303-0019 & $1280$~{\AA} & \citet{Paris2011} & $7$ \\
PSO J0402+2452 & $1280$~{\AA} & \citet{Suzuki2006} & $5$ \\
SDSS J0818+1723 & $1280$~{\AA} & \citet{Paris2011} & $5$ \\
SDSS J0836+0054 & $1280$~{\AA} & \citet{Suzuki2006} & $5$ \\
SDSS J0840+5624 & $1280$~{\AA} & \citet{Paris2011} & $5$ \\
SDSS J0842+1218 & $1280$~{\AA} & \citet{Paris2011} & $5$ \\
SDSS J0927+2001 & $1280$~{\AA} & \citet{Paris2011} & $5$ \\
SDSS J1030+0524 & $1280$~{\AA} & \citet{Paris2011} & $5$ \\
SDSS J1137+3549 & $1280$~{\AA} & \citet{Suzuki2006} & $5$ \\
SDSS J1148+5251 & $1280$~{\AA} & \citet{Paris2011} & $5$ \\
SDSS J1250+3130 & $1280$~{\AA} & \citet{Paris2011} & $5$ \\
SDSS J1306+0359 & $1285$~{\AA} & \citet{Paris2011} & $5$ \\
ULAS J1319+0950 & $1280$~{\AA} & \citet{Suzuki2006} & $5$ \\
SDSS J1335+3533 & $1280$~{\AA} & \citet{Paris2011} & $5$ \\
SDSS J1411+1217 & $1280$~{\AA} & \citet{Paris2011} & $3$ \\
SDSS J1602+4228 & $1280$~{\AA} & \citet{Suzuki2006} & $5$ \\
SDSS J1623+3112 & $1280$~{\AA} & \citet{Paris2011} & $5$ \\
SDSS J1630+4012 & $1280$~{\AA} & \citet{Suzuki2006} & $5$ \\
CFHQS J1641+3755 & $1280$~{\AA} & \citet{Paris2011} & $3$ \\
SDSS J2054-0005 & $1280$~{\AA} & \citet{Paris2011} & $5$ \\
CFHQS J2229+1457 & $1250$~{\AA} & \citet{Paris2011} & $7$ \\
SDSS J2315-0023 & $1280$~{\AA} & \citet{Paris2011} & $5$ \\
CFHQS J2329-0301 & $1230$~{\AA} & \citet{Paris2011} & $5$
\enddata
\tablecomments{The columns show the name of the object, the rest-frame wavelength at which the quasar spectra is normalized to unity, and the set and number of PCS we used to model the quasar continua. }
\end{deluxetable*}

\section{Zones of Remaining Quasars from our Data Set} \label{sec:remaining_zones}

\begin{figure*}[h]
\centering
\includegraphics[width=\textwidth]{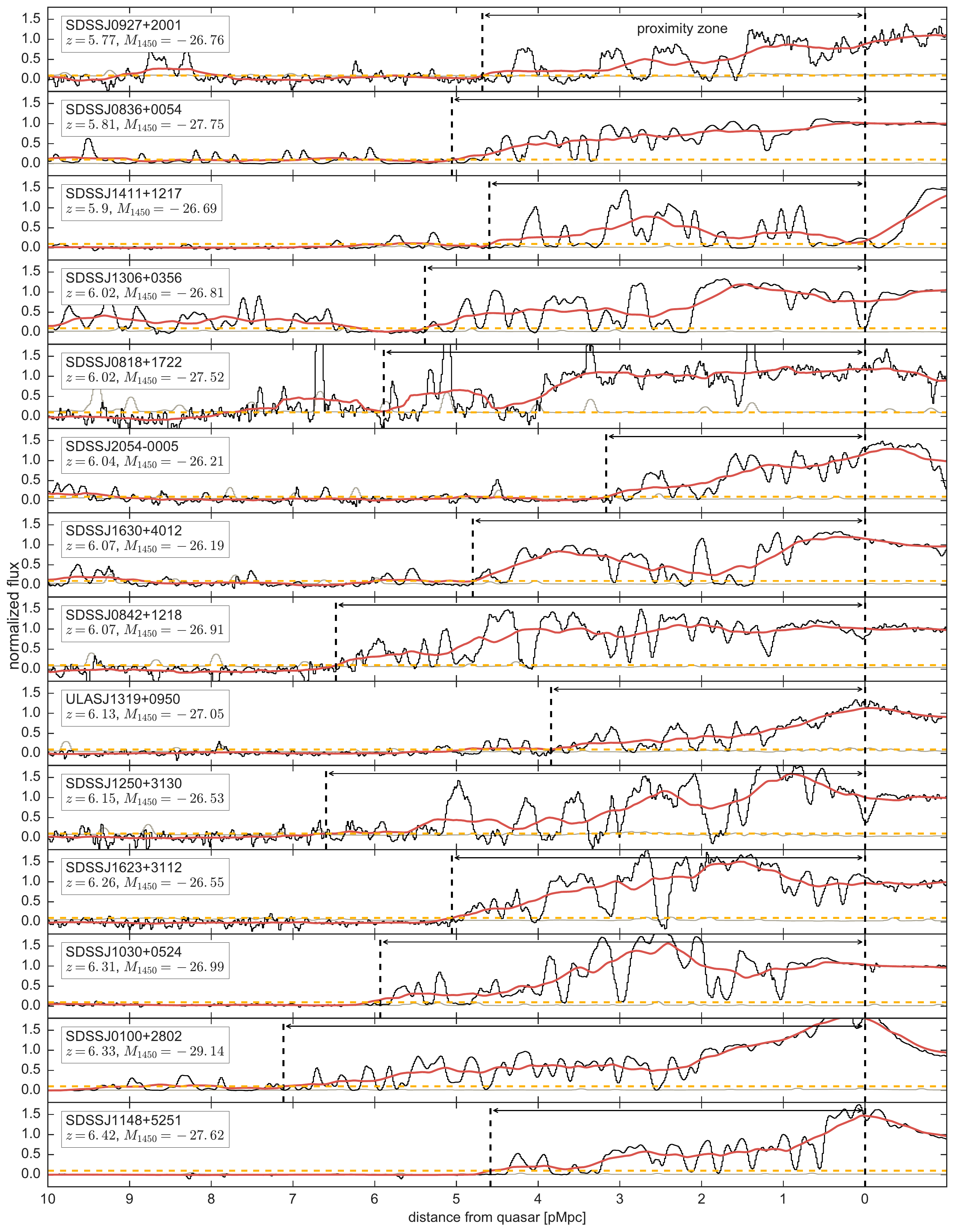}
\caption{Remaining quasar spectra and their proximity zone measurements from our sample sorted by quasar redshift. The spectra (black curves) and their noise vectors (gray curves) are smoothed with a boxcar function of two pixels. The black dashed lines indicate the beginning (location of the quasar) and the end (where the smoothed flux shown as the red curve drops below the $10\%$ level indicated by the yellow dashed lines) of the proximity zones. \label{fig:all_spectra}} 
\end{figure*}

Fig.~\ref{fig:all_spectra} shows the same as Fig.~\ref{fig:9spectra} and Fig.~\ref{fig:9spectra_faint} for the remaining quasars in our data sample sorted by the redshift of the quasar. 

\end{document}